\documentclass{article}

\usepackage{arxiv}

\usepackage{comment}
\usepackage{multirow}
\usepackage{multicol}
\usepackage[utf8]{inputenc}
\usepackage[english]{babel}
\usepackage{graphicx}
\usepackage{amsmath}
\usepackage{amsthm}
\usepackage{amsfonts}
\usepackage{amssymb}
\usepackage{float}
\usepackage{enumerate}
\usepackage{hyperref}
\usepackage{framed}

\title{Onset Detection: A New approach to QBH System}
\author{
  Ritwik Bhaduri\\
  Masters of Statistics\\
  Indian Statistical Institute, Kolkata\\
  \texttt{ritwik.bhaduri@gmail.com}\\ 
   \And
  Soham Bonnerjee\\
  Masters of Statistics\\
  Indian Statistical Institute, Kolkata\\
  \texttt{sohambonnerjee01@gmail.com}\\ 
  \And
  Subhrajyoty Roy\thanks{All authors contributed equally. Ordering is determined in dictionary order of author names.}\\
  Masters of Statistics\\
  Indian Statistical Institute, Kolkata\\
  \texttt{roysubhra98@gmail.com}\\
}

\begin{document}

\maketitle
\begin{abstract}
   Query by Humming (QBH) is a system to provide a user with the song(s) which the user hums to the system. Current QBH method requires the extraction of onset and pitch information in order to track similarity with various versions of different songs. However, we here focus on detecting precise onsets only and use them to build a QBH system which is better than existing methods in terms of speed and memory and empirically in terms of accuracy. We also provide statistical analogy for onset detection functions and provide a measure of error in our algorithm.
\end{abstract}

\keywords{Onset Detection \and STFT \and QBH \and Subset matching \and Correlative matching}

\section{Introduction} 
\qquad Imagine yourself going to a friend's party and hearing some song which you have heard for the first time. You liked the melody, and your subconscious mind picked up the tune of the song, although you cannot seem to recall any of the lyrics of the song later. We, in this report, provide a way out for you to search the song. You just need to hum it, and the algorithms described in the report will do the job for you.

\qquad This problem is known popularly as "Query by humming" (QBH). Conventional approaches to solve this problem proceed by extracting the pitches and notes form the hummed song. However our approach uses the rhythm of the input song rather than the actual notes. This means even if you are not that great a singer you can search your song decently as long as you keep the rhythm more or less correct.

\qquad The entire process involves three main steps:
\begin{enumerate}
    \item \textbf{Building a database of songs.} In our method we don’t need to store the entire songs. We only need to store the onsets of the first line of the song which is very memory efficient compared to storing the entire song in midi or mp3 format.
    \item \textbf{Extracting the onsets from the hummed song.} This is the main step of our algorithm and the performance of the entire system is dependent mainly on how well we can detect the onsets in the input. Onset is the instant where the notes are hit in a hummed song. The formal definition of onset will be provided in the next section.
    \item \textbf{Computing the similarity between a database reference and a user’s query} using certain measurements and produce the final ranked candidates sorted by similarities.
\end{enumerate}

\section{Concepts of Digital Audio}
\qquad In this section, we briefly give an overview of how a signal (such as sound) is captured through a digital system and how it is represented in digital domain. We also discuss about some tools and techniques which would help in solving our problem.

\subsection{Sampling and Quantization}
\qquad We all know that sound is a longitudinal wave and we perceive sound by a continuous and instantaneous changes in air pressure. However, since we can only use a finite amount of memory to store an sound, we cannot observe the amount of air pressure at infinitesimally close points. However, if we only observe the air pressure at only certain points of time, then we are losing a lot of information about that sound, which might affect the quality of perception of that sound wave. 

A simple getaway from this dilemma is to compromise at both end, by choosing a fixed time interval beforehand, and then obtain the values of the air pressure at times maintaining that interval. This interval is chosen so small that during that interval time, the change is air pressure can be assumed to be negligible for hearing. This process is called \textbf{Sampling}. Generally, we use 44100 samples in 1 second, thereby resulting in the time interval of $1/44100 \approx 0.00002267573$ seconds.

\begin{figure}
    \centering
    \includegraphics{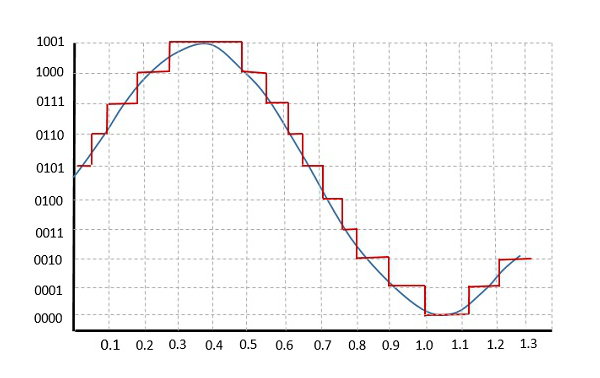}
    \caption{An illustration of Sampling and Quantization with sampling frequency 10 and bitrate of 4}
    \label{fig:samp-quant}
\end{figure}

However the observation of the air pressure is highly unlikely to be a rational number, thereby again demanding for a storage of infinite size to store a single sample. Therefore, we need to approximate the observation using some rational number. More specifically, we divide the whole range of observations into finitely many bins, and approximate the observation by the closest boundary of the bins. The number of such boundaries are generally in some exponent of 2, due to computational ease of storing the signal, and the exponent is called \textbf{bitrate} of the signal. This whole process of binning is called \textbf{Quantization}. An illustration of the process is shown in Figure \ref{fig:samp-quant}.

\subsection{Attack, Onset, Transience, Decay}

\begin{figure}
    \centering
    \includegraphics{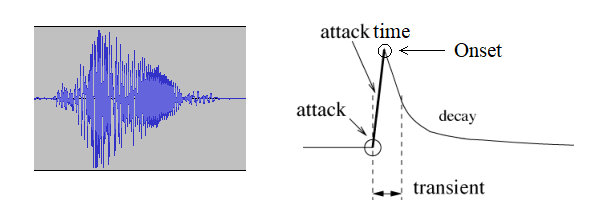}
    \caption{A snapshot of digitally sampled audio\qquad
    Onset, Attack, Decay, Transience in amplitude envelope}
    \label{fig:onset-defn}
\end{figure}

A sound wave or signal has many different attributes which are of interest in the domain of sound processing. The most basic two such attribute which is commonly used in layman terms are \textbf{pitch} and \textbf{beats}, the first of which contains the information about the frequency and amplitude of the notes played in the sound (consider a song, for example), while the second captures the timing information of when a particular note is played. However, this report primarily focuses on the second attribute, hence, we formalize the concepts of \textbf{beats} and some additional concepts which we will be using throughout the report. Figure \ref{fig:onset-defn} shows different attributes related to beat information.

\begin{enumerate}
    \item \textbf{Amplitude Envelope} represents the curve joining the outermost points of the digitally sampled audio. In simplistic terms, it is a smooth curve which represents the general movement of the amplitude at different timepoints in the song.\footnote{We shall be using the word "song" and digitally sampled audio interchangeably for convenience.}
    
    \item \textbf{Attack} is the time interval within which the amplitude envelope steadily increases.
    
    \item \textbf{Decay} is the time interval within which the amplitude envelope steadily decreases. In mathematical terms, during this time, the amplitude envelope can be modelled reasonably well using an exponential decay function $e^{-\lambda t}$, where $\lambda$ is the decaying rate.
    
    \item \textbf{Transient} is a concept which is difficult to comprehend in precise terms. It refers to the time interval for which the sound of the note is dominantly perceivable. Hence, it consists of all of attack time and some time of decay.
    
    \item The \textbf{onset} is the single instant chosen to represent this peculiar change in amplitude envelope. This is usually represented by the starting point of attack time interval. Also, sometimes it is denoted by the peak of the amplitude envelope, i.e. the end of the attack time interval, for ease of analysis. Here, we shall adopt to this second definition. Note that, both of these definitions are equivalent when the attack time interval is assumed to be of length 0, or considerably negligible for the subsequent analysis.
    
    \item \textbf{Offset} is just the opposite of an onset. It may sometimes happens in an audio that there is an onset and the higher value of amplitude envelope remains for a split second before it decreases rapidly. In such case, the decay rate is very high and the endpoint of decaying time interval is called an offset. For instance, such onset and offset can be perceived very clearly for the honking sound of car horn.
    
    \item \textbf{Beat} of a song is the length of the largest time interval for which the time between any two consecutive onsets is an integral multiple of that. In other terms, beat is the largest unit of time based on which, if the time of onsets are measured, all such measured time in \textbf{beat} units will be integers.
\end{enumerate}

\subsection{Discrete Fourier Transform}
\qquad Based on the discussion of Sampling and Quantization, it is clear that digital audio is denoted by a finite sequence or equivalently a vector of numbers, $\textbf{x} = x[0], x[1], \dots x[N-1]$, where $N$ is the number of samples taken in the whole audio. For mathematical convenience, we allow this elements of the sequence $x[i]$ to be any complex number, although in practice we only hear the real components of the sequence.

For complex vectors of length $N$, the euclidean inner product between two vectors $\textbf{x}$ and $\textbf{y}$ such that;
    $$\langle \textbf{x},\textbf{y} \rangle = \sum_{k=0}^{N-1} x[k]\bar{y}[k]$$
    where $\bar{y}[k]$ denotes the complex conjugate of the complex number $y[k]$.
The associated norm of the complex vector $\textbf{x}$ is given by;
    $$\Vert \textbf{x}\Vert = \sum_{k=0}^{N-1} \vert x[k]\vert ^2$$
The pure digital tones of order $N$ of frequency $f$ is the following complex $N$ length vector;
$$v_{f} = \dfrac{1}{\sqrt{N}}\left(1, e^{2\pi i f/N}, e^{4\pi i f/N}, e^{6\pi i f/N}, \dots e^{2\pi i f(N-1)/N}\right)$$
where $i$ is a complex root of the equation $x^2 + 1 = 0$. To mathematician's society, this vector is also known as normalized complex exponentials.
The complete set of all possible pure tones, i.e. the set  $F_N = \left\{v_0, v_1, \dots v_{N-1}\right\}$ is called $N$-point Fourier basis. Note that, these basis is a orthonormal basis with respect to the complex inner product defined above.

The Discrete Fourier Transform (DFT) of a complex valued signal or complex vector $\textbf{x}$ of length $N$ is another complex vector $\textbf{y}$ of length $N$ such that;
$$y[k] = \langle \textbf{x}, v_k \rangle = \dfrac{1}{\sqrt{N}}\sum_{n=0}^{N-1} x[n] e^{-2\pi i kn/N}$$
for $k = 0, 1, \dots (N-1)$.
    
As previously mentioned, the set $F_N$ forms a basis of the vector space of all length $N$ complex vectors. Therefore, we can simply express any signal $\textbf{x}$ (represented by $N$ length complex vectors) as a linear combination of pure digital tones. Let;
$$\textbf{x} = \alpha_1 v_1 + \alpha_2 v_2 + \dots \alpha_Nv_N$$

Hence, performing the inner product of $\textbf{x}$ with any pure tone, $v_f$, we obtain, $\langle \textbf{x}, v_f\rangle = \alpha_f$, by properties of orthonormality. Therefore, the Discrete Fourier Transform of $\textbf{x}$ evaluated at $k$, is actually the coefficient of the pure tone of discrete frequency $k$ (also called $k$-th frequency bin) present in the given signal $\textbf{x}$. In this sense, DFT computes the composition of different pure tones in the signal. Also, if one is solely interested in finding the amount of pure tones present in the song only, he (she) can ignore the multiplicative factor of $1/\sqrt{N}$ in the DFT formula, which leads to loss of some mathematical properties. Using this interpretation of DFT, along with orthonormality of Fourier basis, it is easy to see that;
$$\Vert \textbf{y}\Vert^2 = \Vert \textbf{x} \Vert^2$$
where $\textbf{y}$ is the DFT of the signal $\textbf{x}$. This result is popularly known as \textbf{Parseval's} formula.

Now, for a song, sampling frequency would generally be higher, thereby resulting the size of $N$ to the pretty big. In such case, it is not computationally a good idea to perform DFT over the whole complex vector of size $N$. As noted before, the idea of DFT is to bring out the composition of pure tones in the song, and hence we are only interested in a local phenomenon, i.e. which note is dominantly playing right now, and for that, a distant part of the song has very little effect. Therefore, in practice, we perform DFT over a range of moving window, comprising of Short amount of time. This process is called \textbf{Short Time Discrete Fourier Transform} (STDFT). The STDFT of a signal $\textbf{x}$ at a position $n$ is given by;
$$X_k(n) = \sum_{m= -w/2}^{w/2}x[n+m]e^{-2\pi i km/w}$$
where $w$ is the window length. This $X_k(n)$ is called Short Time Fourier coefficient of discrete frequency $k$ at position $n$.

\section{Onset Detection Algorithms}

\qquad To detect an \textit{onset}, we need to first have a decision algorithm in place, based on the \textit{discrete-time} waveform or \textit{data set} that we have. Mathematically, we assume that the data set consisting $\{ x[0] , x[1] , \cdots, \}$ is available, where $x[i]$ denote the value of the signal (or noise) at time point $t_i=\frac{i}{S_f}$, $S_f$ being the \textit{Sampling Frequency}, i.e the number of samples collected in $1$ second.

\begin{figure}
    \centering
    \includegraphics{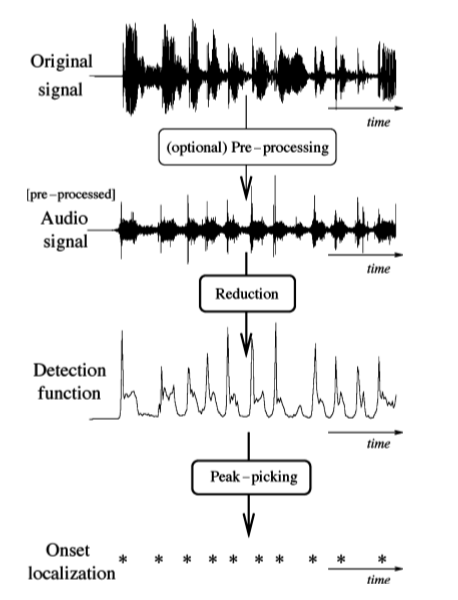}
    \caption{Flowchart of Onset Detection Algorithm}
    \label{fig:onset-flowchart}
\end{figure}

The idea of detecting onsets can be summarized in three main steps:

\begin{enumerate}
    \item \textbf{Preprocessing:} It allows the raw vector of signal to be transformed slightly in order to improve the performance of the subsequent analysis. This step is optional and highly depends on the type of signal you are analyzing.
    \item \textbf{Detection Function:} A reduction is done of the preprocessed signal through a detection function. A detection function is a statistic which sufficiently reduce the data in a more compact form keeping the necessary information about the presence or the strength of the signal in a local neighbourhood. A detection function is applied to the signal through a sliding window (or moving window), allowing only the neighbourhood signal to be summarized. For example, let $\omega$ is the size of the sliding window, and let $T(\cdot)$ be the statistic to be applied. Then, this detection function computes the value of the statistic for each of the moving window;
    $$T[n] = T(x[n], x[n+1], \dots x[n+\omega - 1]) \qquad \forall n = 0,1,2, \dots (N-\omega)$$
    The statistic $T(\cdot)$ is chosen in a way so that the onset at time $t_n$ results in a comparatively higher value of $T[n]$ rather than its other values.
    \item \textbf{Peak Detection:} The resulting detection function is desired to produce local maximums at the time of the true onsets. Therefore, a peak detection algorithm is run at the end to identify the peaks in the vector of detection function. The times corresponding to these peaks are finally identified as possible onsets.
\end{enumerate}

Above discussion has been summarized in a flow chart as shown in Figure \ref{fig:onset-flowchart}. In this section, we describe three different detection functions, while the algorithm for peak detection is described in next section.

\subsection{Energy Detector}
\qquad At first, we consider the detection of presence of a signal in the $n^\text{th}$ time-point. The corresponding detection problem can be modelled as :
\begin{align*}
 H_0: x[n]=w[n]  \hspace{2cm}  n=0,1,\cdots, \omega-1\\ 
 H_1: x[n]=s[n]+w[n] \hspace{1cm} n=0,1,\cdots, \omega-1 
 \end{align*}
where $s[n]$ is deterministic and \textit{completely unknown}, and $w[n]$ is WGN (White Gaussian Noise) with variance $\sigma^2$. 

A GLRT (Generalized Likelihood Ratio Test) would reject $H_0$ in favour of $H_1$ if 
\begin{align}
    \frac{p(\textbf{x};\hat{s}[0], \cdots \hat{s}[\omega-1], H_1) }{p(\textbf{x};H_0)} > \gamma
    \label{eqn1}
\end{align}
where $\hat{s}[n]$ is the MLE under $H_1$. To determine the MLE we maximize the likelihood function:
\begin{align*}
  p(\textbf{x};\hat{s}[0], \cdots \hat{s}[\omega-1], H_1)=\frac{1}{(2\pi\sigma^2)^\frac{\omega}{2}}\exp\left[-\frac{1}{2\sigma^2}\displaystyle\sum_{n=0}^{\omega-1}(x[n]-s[n])^2\right] 
\end{align*}
over the signal samples. Clearly, the MLE is : $\hat{s}[n]=x[n]$. Thus from (\ref{eqn1}), we have:
\begin{align*}
    \dfrac{\frac{1}{(2\pi\sigma^2)^\frac{\omega}{2}}}{\frac{1}{(2\pi\sigma^2)^\frac{\omega}{2}}\exp \left(-\frac{1}{2\sigma^2}\displaystyle\sum_{n=0}^{\omega-1}x^2[n]\right)} > \gamma
\end{align*}
Taking logarithm produces
\begin{align*}
    \frac{1}{2\sigma^2}\displaystyle\sum_{n=0}^{\omega-1}x^2[n]> \ln \gamma
\end{align*}
Equivalently, we reject $H_0$ in favour of $H_1$ if
\begin{align}
   T(\textbf{x})= \displaystyle\sum_{n=0}^{\omega-1}x^2[n] > \gamma'
   \label{eqn2}
\end{align}
This detector computes the \textit{local energy} in the received data and compares it to a threshold. Hence this statistic is known as \textit{Energy Detector}.

\subsection{Spectral Dissimilarity}
\begin{figure}
    \centering
    \includegraphics{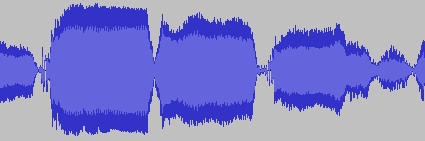}
    \caption{Effect of Meend on Amplitude Envelope}
    \label{fig:onset_blob}
\end{figure}

\qquad There are two drawbacks with Local Energy Detector as mentioned in previous section.
\begin{enumerate}
    \item Firstly, the local energy detector computes the sum of squared values of the signal. However, when we speak or hum, it is evident that a lot of the energy is actually contained in the pure tones of lower frequency (about 80-200 Hz). Hence, the energy detector also fluctuates a lot because of changes in noise at higher frequency level, and will return some false peaks.
    \item Secondly, the energy detector outputs a detection function which returns a peak at the location of true onset, only if the time interval of attack is negligible. However, in specific, for Rabindra Sangeets, there is abundance of \textbf{Meends} \footnote{In its simplest form, a meend is a smooth glide from one note to another.} which challenges this hypothesis. In such case, the amplitude envelope rises at the time of the attack, reaches the highest point, determining the onset, and stays at high level for a nonnegligible period of time, and finally decays. Figure \ref{fig:onset_blob} shows one such example.
\end{enumerate}

For the first problem, we make effective use of \textbf{Parseval's Formula} to obtain that the local energy computed would be same as the energy computed for each discrete frequency. Since, the fundamental frequency of humming usually ranges from 80 to 200 Hz, and usually at 3 or 4 harmonics are present, we can safely assume that the frequency bins of ranges higher than 1KHz, generally contains no information about the hummed song and hence can be discarded via performing STDFT. This process helps us to remove unwanted noises from the signal.

For the second problem, it is evident that the problem with Energy Detector is that it is unable to detect immediate changes in local energy. When an onset appears, it is clear that there is some changes, both in amplitude and in frequencies. Therefore, one way to obtain the change is to consider the difference in magnitude of $k$-th frequency bin, i.e. to consider $\vert X_k(n) \vert - \vert X_k(n-1) \vert$, which is able to capture the immediate change in that frequency bin only. However, presence of an offset also results in such changes in magnitude of the frequency bin, but that would generally be negative. Therefore, we consider the following statistic,
$$T(n) = \sum_{k} \left(\vert X_k(n) \vert - \vert X_k(n-1) \vert\right) \textbf{1}_{\left\{\vert X_k(n) \vert > \vert X_k(n-1) \vert\right\}}$$
where $\textbf{1}_{A}$ denotes the indicator function of event $A$. The sum in the above statistic extends over all discrete frequency bins for which the real frequency is at most 1KHz, summarizing the changes in those discrete frequencies via $L_1$ norm.
The above test statistic is called \textbf{Spectral Dissimilarity} and is used to detect onsets of the hummed song.

\subsection{Dominant Spectral Dissimilarity}

\qquad We can assume our signal is sinusoid in nature, and hence our onset detection problem can be equivalently stated as a problem of detection of a sinusoid in \textit{WGN}. The general detection problem is:
\begin{align*}
H_0: x[n]&=w[n] && n=0,1,\cdots, \omega-1\\\\ 
H_1: x[n]&= \smash{\left\{\begin{array}{@{}l@{}}
           w[n] \\[\jot] 
           A\cos(2\pi f_0n+\phi)+w[n] \\[\jot] 
           \end{array}\right.} && n=0,1,\cdots, n_0-1, n_0+M, \cdots, \omega-1 \\
  & && n=n_0,n_0+1,\cdots, n_0+M-1
 \end{align*}
where $w[n]$ is WGN with known variance $\sigma^2$, and $A$, $f_0$, $\phi$ are \textit{Amplitude}, \textit{Frequency}, and \textit{Phase} respectively. These parameters might be unknown. $n_0$ is the \textit{Lag-time} and sinusoid is non-zero over the interval $[n_0, n_0+M-1]$ where $M$ is the signal length. We assume that $n_0=0$, and then the observational interval is just the signal interval , i.e $[0,\omega-1]=[0,M-1]$. Then the detection problem is:
\begin{align*}
 H_0: x[n]=w[n]  \hspace{4.05cm}  n=0,1,\cdots, \omega-1\\ 
 H_1: x[n]= A\cos(2\pi f_0n+\phi)+w[n] \hspace{1cm} n=0,1,\cdots, \omega-1 
 \end{align*}
 We will consider the following two cases:
 \begin{enumerate}
     \item $A$, $\phi$ unknown.
     \item $A$, $\phi$, $f_0$ unknown.
 \end{enumerate}
\subsubsection{Amplitude and Phase Unknown}
When $A$ and $\phi$ are unknown, then we must assume that $A>0$. Otherwise, for two different sets of $(A, \phi)$ we will get the same signal, for example if $A=1$, $\phi=0$ we have the signal$=\cos(2\pi f_0n)$, and for $A=-1$, $\phi=\pi$, we have signal=$\cos(2\pi f_0n+\pi)=\cos(2\pi f_0n)$. Hence, the parameters will not be \textit{identifiable}.

Assuming $A>0$, the GLRT rejects $H_0$ in favour of $H_1$ if:
\begin{align*}
    \frac{p(\textbf{x};\Hat{A}, \Hat{\phi}, H_1) }{p(\textbf{x};H_0)} > \gamma
\end{align*}
where $\Hat{A}, \Hat{\phi}$ are the corresponding MLE's. Equivalently, we reject $H_0$ in favour of $H_1$ if:
\begin{align}
   L_G(\textbf{x})= \dfrac{\frac{1}{(2\pi\sigma^2)^\frac{\omega}{2}}\exp \left[-\frac{1}{2\sigma^2}\displaystyle\sum_{n=0}^{\omega-1}\left(x[n]-\Hat{A}\cos(2\pi f_0n+\Hat{\phi})\right)^2\right]}{\frac{1}{(2\pi\sigma^2)^\frac{\omega}{2}}\exp \left[-\frac{1}{2\sigma^2}\displaystyle\sum_{n=0}^{\omega-1}x^2[n]\right]} > \gamma
\end{align}
For sufficiently large $\omega$, we have, approximately:
\begin{align}
    \Hat{A}=\sqrt{\Hat{{\alpha_1}}^2+\Hat{{\alpha_2}}^2}\\
    \Hat{\phi}=\arctan(-\frac{\Hat{\alpha_2}}{\Hat{\alpha_1}})
\end{align}
where,
\begin{align}
    \Hat{\alpha_1}=\frac{2}{\omega}\displaystyle\sum_{n=0}^{\omega-1}x[n]\cos{2\pi f_0n}\\
    \Hat{\alpha_2}=\frac{2}{\omega}\displaystyle\sum_{n=0}^{\omega-1}x[n]\sin{2\pi f_0n}
\end{align}
Now,
\begin{align*}
    \ln L_G(\textbf{x})=-\frac{1}{2\sigma^2}\left[\displaystyle\sum_{n=0}^{\omega-1}-2x[n]\Hat{A}\cos(2\pi f_0n+\Hat{\phi})+\displaystyle\sum_{n=0}^{\omega-1}\Hat{A}^2\cos^2(2\pi f_0n+\Hat{\phi}) \right]
\end{align*}
Noting that we can write $\Hat{\alpha_1}=\Hat{A}\cos\Hat{\phi}$, and $\Hat{\alpha_2}=-\Hat{A}\sin\Hat{\phi}$, we have:
\begin{align*}
  \displaystyle\sum_{n=0}^{\omega-1}x[n]\Hat{A}\cos(2\pi f_0n+\Hat{\phi})&= \displaystyle\sum_{n=0}^{\omega-1}x[n]\cos(2\pi f_0n)\Hat{A}\cos{\Hat{\phi}}-\displaystyle\sum_{n=0}^{\omega-1}x[n]\sin(2\pi f_0n)\Hat{A}\sin{\Hat{\phi}}\\
  &=\frac{\omega}{2}(\Hat{{\alpha_1}}^2+\Hat{{\alpha_2}}^2)\\
\end{align*}
Using $\displaystyle\sum_{n=0}^{\omega-1}\cos^2(2\pi f_0n+\Hat{\phi})\approx\frac{\omega}{2}$, we have
\begin{align*}
    \ln L_G(\textbf{x})&=-\frac{1}{2\sigma^2}\left[-2\frac{\omega}{2}(\Hat{{\alpha_1}}^2+\Hat{{\alpha_2}}^2)+\frac{\omega}{2}\Hat{A}^2   \right]\\
    &= -\frac{1}{2\sigma^2}\left[ -\frac{\omega}{2}(\Hat{{\alpha_1}}^2+\Hat{{\alpha_2}}^2)   \right]\\
    &=\frac{\omega}{4\sigma^2}(\Hat{{\alpha_1}}^2+\Hat{{\alpha_2}}^2)
\end{align*}
So, we reject $H_0$ in favour of $H_1$ if:
\[ \frac{\omega}{4\sigma^2}(\Hat{{\alpha_1}}^2+\Hat{{\alpha_2}}^2)>\ln \gamma \]
But, 
\begin{align*}
\Hat{{\alpha_1}}^2+\Hat{{\alpha_2}}^2&=(\dfrac{2}{\omega})^2\left[\left(\displaystyle\sum_{n=0}^{\omega-1}x[n]\cos{2\pi f_0n}\right)^2 + \left(\displaystyle\sum_{n=0}^{\omega-1}x[n]\sin {2\pi f_0n}\right)^2 \right]\\
&=\dfrac{4}{\omega}\dfrac{1}{\omega}\left|\displaystyle\sum_{n=0}^{\omega-1}x[n]\exp(-i2\pi f_0n)\right|^2\\
&=\dfrac{4}{\omega}I(f_0)\\
\end{align*}
where $i^2=-1$, and $I(f_0)$ is the \textit{periodogram} at $f=f_0$. 

Hence, we reject $H_0$ in favour of $H_1$ if
\begin{align}
 I(f_0)>\sigma^2\ln \gamma=\gamma'
 \label{EQN}
\end{align}

Note that, testing with large values of $I(f_0)$ is equivalent to considering large values of the magnitude of the Discrete Fourier Coefficient of the signal $x$ at the known frequency $f_0$.

\subsubsection{Amplitude, Phase and Frequency Unknown}
In this case, the GLRT rejects $H_0$ if:
\begin{align*}
    \frac{p(\textbf{x};\Hat{A}, \Hat{\phi}, \Hat{f_0}, H_1) }{p(\textbf{x};H_0)} > \gamma
\end{align*}
which is equivalent to:
\begin{align*}
    \frac{\displaystyle\max_{f_0}p(\textbf{x};\Hat{A}, \Hat{\phi}, f_0, H_1) }{p(\textbf{x};H_0)} > \gamma
\end{align*}
Note that under $H_0$, $p(\textbf{x})$ does not depend on $f_0$, and is non-negative; hence we can write the rejection region as:
\begin{align*}
    \displaystyle\max_{f_0}\frac{p(\textbf{x};\Hat{A}, \Hat{\phi}, f_0, H_1) }{p(\textbf{x};H_0)} > \gamma
\end{align*}
Using the monotonicity of logarithmic function, this rejection region can be represented as:
\begin{align*}
   \ln \displaystyle\max_{f_0}\frac{p(\textbf{x};\Hat{A}, \Hat{\phi}, f_0, H_1) }{p(\textbf{x};H_0)} > \ln \gamma
   \Longleftrightarrow  \displaystyle\max_{f_0}\ln\frac{p(\textbf{x};\Hat{A}, \Hat{\phi}, f_0, H_1) }{p(\textbf{x};H_0)} > \ln \gamma \\
\end{align*}
Now, from equation \ref{EQN}, we have:
\begin{align*}
\ln \frac{p(\textbf{x};\Hat{A}, \Hat{\phi}, f_0, H_1) }{p(\textbf{x};H_0)}=\frac{I(f_0)}{\sigma^2}
\end{align*}
Hence, finally, we reject $H_0$ in favour of $H_1$ if:
\begin{align}
    \displaystyle\max_{f_0}I(f_0)>\sigma^2\ln\gamma=\gamma'
\end{align}
Hence, the detector accepts the presence of sinusoidal wave if the peak value of the periodogram exceeds a threshold.

Here, testing with large values of $\max_{f_0} I(f_0)$ is equivalent to considering maximum of the magnitude of the Discrete Fourier Coefficients of the signal $x$ and comparing this maximum value to a threshold.

This motivates the construction of \textit{Dominant Spectral Dissimilarity} Detection Function. As noted earlier, it is required that the detection function observes the changes in amplitude envelope rather than it absolute value. Combining the idea of maximum periodogram along with Spectral dissimilarity, we consider the following detection function:

$$T[n] = \left(\max_k \vert X_k(n)\vert^2 - \max_k\vert X_k(n-1)\vert^2\right)\times \textbf{1}_{\left\{\max_k \vert X_k(n)\vert > \max_k\vert X_k(n-1)\vert\right\}}$$

In contrast of Spectral Dissimilarity, where we consider changes in all frequency content, here, we consider changes in only the dominant frequency content. This allows us to nullify the effect of subtle changes in frequency spectrum during random noises, thereby conveying only the relevant information.

\subsection{Peak Detection}
\qquad After obtaining the values of the reduction function $T(\cdot)$ at some regular interval, we output the possible peaks of the function as our output of onset. Before proceeding with the description of peak detection algorithm we have used, we describe some important features of a Peak.
\begin{enumerate}
    \item Due to randomness in noise, it might happen that the detection function shows a peak at a location where there is only noise. But from the desirable properties of detection function, it is evident that such peaks would be of relatively smaller height than a peak where onset has occurred. Therefore, we must choose a threshold parameter so that the peaks below that threshold parameter is completely ignored. Note that, this threshold parameter should be adaptive to the changes in detection function.
    \item As the name suggest, a peak should be of a higher value that its neighbouring values of detection function. Therefore, to assign a point of time as peak, we must compare its value of detection function with that of its neighbours, and such neighbourhood has to be predetermined.
    \item Finally, two peaks should not be too close to each other. It is a commonly known fact that two consecutive sounds must be at least $1/10$-th of a second apart in time to be heard as distinguished sounds. Therefore, a reasonable peak detection algorithm should merge two onsets into a single one if they are less than 0.1 second apart.
\end{enumerate}

Keeping above desirable properties in mind, we perform the peak detection in the most simplistic way possible. The algorithm can be described in the following points:

\begin{enumerate}
    \item \textbf{Input:} The value of the detection function at timepoints $t_0, t_1, \dots t_{N-\omega}$, A hopsize $h$, an integer vector $a_1, a_2, \dots a_r$ defining the neighbouring points, and a threshold criterion.
    \item For each timepoint $t_k$ in $\left\{t_0, t_{h}, t_{2h}, \dots t_{\lfloor \frac{N-\omega}{h}\rfloor}\right\}$:
    \begin{enumerate}
        \item Obtain the value of the detection function at discrete time $k$ i.e. $T[k]$.
        \item Check whether $T[k]$ meets the given threshold criterion. If yes, go to next step. Else, go to next iteration of the loop.
        \item Check whether $T[k] > T[k+a_i]$ for all $i = 1,2,\dots r$. If yes, go to next step. Else, go to next iteration of the loop.
        \item Output $t_k$ as a time of onset. Then move to the iteration with $t_{k'}$ such that, $k' = \min\left\{i : t_{i} - t_{k} > 0.1\right\}$.
    \end{enumerate}
\end{enumerate}

Usually, hopsize is taken to be 1, while the integer vector defining the neighbouring points is usually taken as $(-r), (-r+1), \dots (-1), 1, 2, \dots r$. However, for some use cases, it might help to allow these neighbouring points to be asymmetric about 0. Also, the threshold criterion is effectively used in order to prevent the detection of false peaks. There are two simple yet effective choices for threshold criterion,

\begin{enumerate}
    \item One is to consider the mean of the detection function as the threshold. In such case, any peak above the mean value of the detection function (or some suitable constant multiplied with it) will be detected as possible onset.
    \item Another possible choice of the threshold is to consider the 3rd quartile of the detection function. This is particularly useful when the detection function does not generally have an upper bound.
\end{enumerate}

It is found that increasing the hopsize results in decreasing the number of false positives, while an increment in the number of false negatives and vice versa. Similarly, increase in $r$, the number of neighbouring values to compare with, greatly reduces the number of false positives, while some true onsets can be missed, if it is "very" close to another true onset with a higher peak value. Keeping our goal in mind and with the searching techniques proposed in next section, having a few false positives do not provide us a bad situation, while missing some true onsets can greatly decrease the accuracy of the whole system.

\subsection{Power of the Detection Algorithms}
\qquad In this section, we discuss some simple probability bounds which can help us in obtaining the power of the detection algorithms. This will allow us to obtain an expression of the standard error made by the algorithm for detecting an onset. 

We consider the following mathematical setup. We observe a sequence of numbers, $x[0], x[1], \dots x[N-1]$, which constitutes the signal. $x[k]$ denotes the value of the signal at timepoint $t_k = k/S_f$, where $S_f$ is the sampling frequency. There is a single onset present at the timepoint $t^* = k^*/S_f$, before which the signal only constitutes of random Gaussian noise with mean $0$ and constant variance $\sigma^2$. However, after this onset, the signal is modelled by a decaying sinusoidal. Therefore,

\begin{align*}
 x[k] & \sim N(0, \sigma^2) &&  k=0,1,\cdots, k^*-1\\ 
 x[k] & \sim N(A\exp{(\lambda(k-k^*)/S_f)}\cos{\left(2\pi f_0(k - k^*)/S_f\right)}, \sigma^2) && k=k^*,k^* + 1,\cdots, N-1 
 \end{align*}

where $A$ is the initial amplitude of the signal, $\lambda$ is the decaying parameter and $f_0$ is the frequency of the note played at the time of the onset. Note that, these quantities are usually unknown, but these parameters are useful in order to study the properties of the power function. To obtain an expression for the power function, we assume that $T[k]$ is the value of the detection function at $k$ with a moving window of length $\omega$, and the peak detection algorithm compares $r$ neighbours to both sides and uses a hopsize of $h$ samples. Also, let $\alpha$ denote the threshold value used for threshold criterion during peak detection.

\begin{align*}
    P\left(\hat{k}\text{ is an outputted onset}\right) & = P\left(T[\hat{k}]\text{ is a peak}, T[\hat{k}] > \alpha\right)\\
    & = P\left(T[\hat{k}] > \alpha, T[\hat{k}] > T[\hat{k} + nh], n = (-r), (-r+1), \dots (-1), 1, \dots r\right)\\
    & \geq \min\left\{0, \sum_{n=(-r)}^{r}P(T[\hat{k}] > T[\hat{k} + nh]) + P(T[\hat{k}] > \alpha) - 2r\right\}
\end{align*}

The last step follows simply from the use of Boole-Frechet inequality. Note that, if the distribution of the detection function $T[k]$ can be obtained explicitly, then the above lower bound can be computed. Also note that, the probabilities $P(T[\hat{k}] > T[\hat{k} + nh])$ should depend on the difference $(\hat{k}-k^*)$, as specification of the quantity $(k-k^*)$ only allows us to obtain the distribution of $x[k]$. Therefore, we can plot the above lower bound against the deviations $(k-k^*)$, from which it would be evident to obtain the probability of detecting an onset within some specified neighbourhood of the true onset $k^*$. 

On the other hand, to bound the number of false positives, it is enough to provide an upper bound of the probability that $\hat{k}$ is an outputted onset. For this case, we again use another Boole-Frechet inequality.

\begin{align*}
    P\left(\hat{k}\text{ is an outputted onset}\right) & = P\left(T[\hat{k}]\text{ is a peak}, T[\hat{k}] > \alpha\right)\\
    & = P\left(T[\hat{k}] > \alpha, T[\hat{k}] > T[\hat{k} + nh], n = (-r), (-r+1), \dots (-1), 1, \dots r\right)\\
    & \leq P(T[\hat{k}] > T[\hat{k} + rh], T[\hat{k}] > \alpha)
\end{align*}

Now note that, if we have $rh \geq \omega$, the window size, then $T[k]$ and $T[k+rh]$ are independent of each other, as they are obtained from different signal values. Also, if we assume at the time $t_{\hat{k}+ rh}$, there is only noise present in the signal (i.e. $\hat{k} < k^* - rh$), then it follows that at time $t_{\hat{k}}$, there is also only random noise present in the signal. Therefore, $T[\hat{k}]$ and $T[\hat{k}+rh]$ would be independent and identically distributed if $\hat{k}$ is at least $rh$ samples before than the true onset. 

Now, let us consider two i.i.d. continuous random variables $X$ and $Y$, and consider the following chain of equalities;

\begin{align*}
    P(X > Y, X> \alpha) & = P(X > Y, X > \alpha, Y > \alpha) + P(X > Y, X > \alpha, Y \leq \alpha)\\
    & = P(X > Y, Y > \alpha) + P(X > \alpha, Y \leq \alpha)\\
    & = P(X < Y, X > \alpha) + P(X > \alpha) P(Y\leq \alpha)\\
    & = P(X > \alpha) - P(X > Y, X> \alpha) + P(X > \alpha) P(Y\leq \alpha)
\end{align*}

In the last line, we use the fact that $P(X = Y) = 0$, since $X$ and $Y$ are continuous random variables. Therefore, we obtain;
$$P(X > Y, X> \alpha) < \frac{1}{2}\left( P(X>\alpha) (2 - P(X> \alpha)) \right)$$

Applying this, we obtain, when $\hat{k}$ is at least $rh$ samples before than the true onset; 

$$P\left(\hat{k}\text{ is an outputted onset}\right) < \frac{1}{2}\left( P(T[\hat{k}]>\alpha) (2 - P(T[\hat{k}] > \alpha)) \right)$$

which only requires the distribution of the detection function applied on a bunch of random Gaussian noise.

Note that, if $\hat{k}$ appears after $rh$ many samples from the true onset $k^*$, we can use $T[\hat{k}]$ and $T[\hat{k}-rh]$ in the above procedure. However, though they are independent, these are not identically distributed. But, from the desirable properties of the detection function $T[\cdot]$, we know that $\hat{k}-rh$ being closer to the true onset $k^*$, $T[\hat{k}-rh]$ should be stochastically larger than $T[\hat{k}]$, should satisfy the inequality; $P(T[\hat{k}]> T[\hat{k} - rh]) < P(X > Y)$, where both $X$ and $Y$ are independent and identically distributed as $T[\hat{k}]$. Therefore, assuming that the detection function is \textit{reasonable}, i.e. satisfies the above property, then the above bound can be safely used for the timepoints after the true onset value.

\subsubsection{Power of Local Energy Detector}
\qquad Consider the local energy detector, where the detection function at discrete timepoint $k$, based on a window of size $\omega$ is given by;
$$T[k] = \sum_{i=0}^{\omega - 1} (x[k+i])^2$$

We divide the derivation of the distribution of this into three cases as follows:

\begin{enumerate}
    \item If $k+\omega -1 < k^*$, then clearly, the window contains only noises in the signal. Therefore, $T[k]/\sigma^2$ follows a central chi-sqaured distribution with $\omega$ degrees of freedom.
    \item If $k \leq k^* \leq (k + \omega - 1)$, then the window contains the true onset. Therefore, $T[k]/\sigma^2$ follows a non-central chi-sqaured distribution with $\omega$ degrees of freedom, where the noncentrality parameter is given as;
    \begin{align*}
        ncp & = \dfrac{1}{\sigma^2}\sum_{i=k^*}^{(k + \omega - 1)}{A^2 \exp{\left( -2\lambda \dfrac{i-k^*}{S_f}\right)} \cos^2{\left( 2\pi f_0 \dfrac{i-k^*}{S_f} \right)}}\\
        & = \dfrac{S_f}{\sigma^2} \times \dfrac{1}{S_f}\sum_{i=k^*}^{(k + \omega - 1)}{A^2 \exp{\left( -2\lambda \dfrac{i-k^*}{S_f}\right)} \cos^2{\left( 2\pi f_0 \dfrac{i-k^*}{S_f} \right)}}\\
        & \approx \dfrac{S_f}{\sigma^2} \times \int_{0}^{t_{(k - k^* + \omega - 1)}} {A^2 \exp{\left( -2\lambda u\right)} \cos^2{\left( 2\pi f_0 u\right)} du}\\
        & = I(k + \omega - 1)
    \end{align*}
    \item If $k^* < k$, then the true onset appears before the current window, therefore, we observe only a decaying signal in the window. Hence, $T[k]/\sigma^2$ follows a non-central chi-sqaured distribution with $\omega$ degrees of freedom, where the noncentrality parameter is given by $ncp = I(k + \omega - 1) - I(k)$. Note that, this integral, $I(k)$ depends only on the difference between $k$ and $k^*$, not on their absolute values. Also, the integral depends on the \textit{Signal to Noise Ratio} i.e. on $A/\sigma$, rather than their individual values.
\end{enumerate}

On the other hand, the probabilities $P(T[k] > T[k + h])$ can be computed by noting that,

\begin{align*}
    P(T[k] > T[k - h]) & = P(\sum_{i=0}^{\omega - 1}x^2[k+i] > \sum_{i=0}^{\omega - 1}x^2[k+h+i])\\
    & = P(\sum_{i=0}^{h - 1}x^2[k+i] > \sum_{i=\omega}^{\omega + h - 1}x^2[k+i]) \\
    & = P\left(\dfrac{\sum_{i=0}^{h - 1}x^2[k+i]}{\sum_{i=\omega}^{\omega + h - 1}x^2[k+i]} > 1\right)
\end{align*}

In above, the quantities $\sum_{i=0}^{h - 1}x^2[k+i]$ and $\sum_{i=\omega}^{\omega + h - 1}x^2[k+i]$ are independently distributed chi sqaured random variables with same degrees of freedom $h$, however, with different non-centrality parameter. Hence, the distribution of the above random variable (the ratio of chi-sqaured r.v.) is doubly non-central F distribution, and the above probability can be explicitly computed.

\subsubsection{Spectral based Detection}
\qquad To study the effect of Dominant Spectral Dissimilarity, we need to revise our model to incorporate complex valued random variable. In this case, we consider the following description of the model, where the signal is allowed to be represented by a complex valued vector, but we only gets to observe the real part of it.

\begin{align*}
 \Re{(x[k])} & \sim N(0, \sigma^2) && k=0,1,\cdots, k^*-1\\ 
 \Re{(x[k])} & \sim N(A\exp{(\lambda(k-k^*)/S_f)}\cos{\left(2\pi f_0(k - k^*)/S_f\right)}, \sigma^2) && k=k^*,k^* + 1,\cdots, N-1 \\
 \Im{(x[k])} & \sim N(0, \sigma^2) && k = 0,1,\cdots N-1
\end{align*}

We also assume that $\Re{(x[k])}$ and $\Im{(x[k])}$ are independent of each other. Now, consider the signal within a window of length $\omega$ as $x[n], x[n+1], \dots x[n+\omega - 1]$. Denote this $\omega$-length vector by $\textbf{x}$ say. Then, $\textbf{x} \sim CN(\mu, \sigma^2\text{I}, 0)$, i.e. $\textbf{x}$ is a complex valued random vector distributed according to a Complex Normal Distribution with some mean $\mu$, determined from the description of the model, $\sigma^2\text{I}$ as covariance matrix and $0$ as relation matrix. Now, consider the matrix $F_\omega$ where each of columns are pure digital tones of order $\omega$, then it is Hermitian, which respect to the complex inner product defined earlier. Based on this, we obtain the distribution of the DFT of $\textbf{x}$ as;
$$DFT\left(\textbf{x}\right) = F_\omega\textbf{x} \sim CN(F_\omega \mu, \sigma^2 F_\omega F_{\omega}^{H}, 0)$$

From the orthonormality of Fourier basis, we have $DFT\left(\textbf{x}\right) \sim CN(DFT\left(\mu\right), \sigma^2\text{I}, 0)$, which shows that the imaginary part and real part of the Fourier coefficients are independent of each other. Hence, the Short Time Fourier coefficient $X_{k}(n) \sim CN(\mu_k(n), \sigma^2, 0)$, where $\mu_k(n)$ denotes the STDF coefficient of discrete frequency $k$ of the mean signal at time $k$. Therefore, the quantity $\dfrac{2}{\sigma^2}\vert X_k(n)\vert^2$ follows a non-central chi-squared distribution with non-centrality parameter given by $\vert\mu_k(n)\vert^2$.

Note that, the random variables $\dfrac{2}{\sigma^2}\vert X_1(n)\vert^2, \dfrac{2}{\sigma^2}\vert X_2(n)\vert^2, \dots \dfrac{2}{\sigma^2}\vert X_\omega(n)\vert^2$ are independently distributed. However, $\dfrac{2}{\sigma^2}\vert X_k(n)\vert^2$
and $\dfrac{2}{\sigma^2}\vert X_k(n-1)\vert^2$ are not independently distributed, hence the distribution of their differences is difficult to obtain. Also, in dominant spectral dissimilarity, we have the difference of the maximum of such non-central chi-squared distributions, which is again difficult to obtain in form of any known distribution.

\section{Searching Techniques}

\qquad In this project we ultimately aim to provide the song that best matches the hummed song from our database. This involves three main steps:
\begin{enumerate}
    \item Forming a database of songs in a suitable format so that we can the results of our previous algorithms to compare the similarity of the input song and the songs in the database.
    \item Comparing the output of the previous algorithms i.e. the onsets in the hummed song with the songs in the repository.
    \item Returning the songs in the repository with the maximum match with the hummed song.
\end{enumerate}

For completing the first step, the onsets of the song are noted down and are stored in the database. This can be done by hand by music experts or from many online sources. These onsets in the database is in the units of time measure (or beat) of the song, which can be easily obtained from the score (notations) of the song.

Now that we have obtained a set of onsets from detection algorithms we have to identify the subset of onsets that’s are true onsets and those which are just false positives. We strongly recommend the idea of using a detection algorithm which is extremely less prone to false negatives as a consequence, as the searching algorithm cannot use an information about an existing onsets in the song, which was not detected by the detection algorithm. Anyway, to perform this searching, we use an algorithm called \textbf{Correlative matching}. Correlative matching uses another subroutine called \textbf{Subset matching}. Both the algorithms are described below. After we have obtained the refined set of onsets from our output we calculate the correlation between our output and the true onsets. The song(s) with the maximum correlation(s) will be returned as the best match(es).

\subsection{Subset Matching}
\subsubsection{The Problem}
Suppose we have an input song and a song from the repository. We have identified a set of probable onsets by our previous algorithms. Let this set of onsets be called Output\_onsets. Also we have a set of onsets for the song in the database which is identified by experts (or from some other source). This set of onsets is assumed to be the true onset and so, it is called True\_onsets. 

Now, Output\_onsets contains some false positives (Onsets that are absent in the True\_onsets i.e. the onsets that are falsely detected) as well as some false negatives (true onsets which our algorithm has failed to detect). Our task is to identify the false positives and false negatives so that we can the two vectors of onsets.

\subsubsection{Assumption}

Both the songs are assumed to be in the same unit with respect to time. It means that the same verse is present in both the songs and the time taken to sing both the parts is equal. This assumption will be discarded later on in Correlative matching. This assumption allows us to lay Output\_onsets and True\_onsets on top of each other and then find the matching.

We use another subtle assumption to efficiently implement the Subset matching algorithm. We assume that both the Output\_onsets and True\_onsets are sorted in increasing order of time. This assumption is reasonable as all of the Detection algorithm we have considered before detects the peaks of the Detection function in increasing order of time.

\subsubsection{The Algorithm}

The input to the algorithm is;
\begin{enumerate}
    \item \textbf{Output\_onsets:} Onsets of the hummed song (input) which has some false positives as well as some false negatives.
    \item \textbf{True\_onsets:} Onsets of the song in the database that is assumed to be true.
\end{enumerate}

The output of this algorithm is the set of onsets which is a subset of our Output\_onsets which does not contain any false positives. It also outputs a subset of True\_onsets that is correctly detected by the onset detection algorithms.

The algorithm can be described as follows;

\begin{enumerate}
    \item Consider the Output\_onsets and True\_onsets and plot them together in a straight line.
    \item Now for every Output\_onset do the following:
    \begin{enumerate}
        \item Find the True\_onset nearest to the Output\_onset (Call it $Y$).
        \item Find the nearest Output\_onset to $Y$ (Call it $X$).
        \item If $X$ = chosen Output\_onset then label the chosen Output\_onset “True positive”.
        \item Else label the chosen Output\_onset “False positive”.
    \end{enumerate}
    \item For every True\_onset do the following:
    \begin{enumerate}
        \item Find the Output\_onset nearest to the True\_onset (Call it $X$).
        \item Find the nearest True\_onset to $X$ (Call it $Y$).
        \item If $Y$ = chosen True\_onset then label the chosen True\_onset “Correctly detected”.
        \item Else label the chosen True\_onset “False negative“
    \end{enumerate}
    \item Matched entry = Output\_onsets which are not False positives.
    \item Detected onsets = True\_onsets which are not False negatives.
    \item Return False negatives, False positives, Matched entry and Detected onsets.
\end{enumerate}

\begin{figure}
    \centering
    \includegraphics[width = \textwidth]{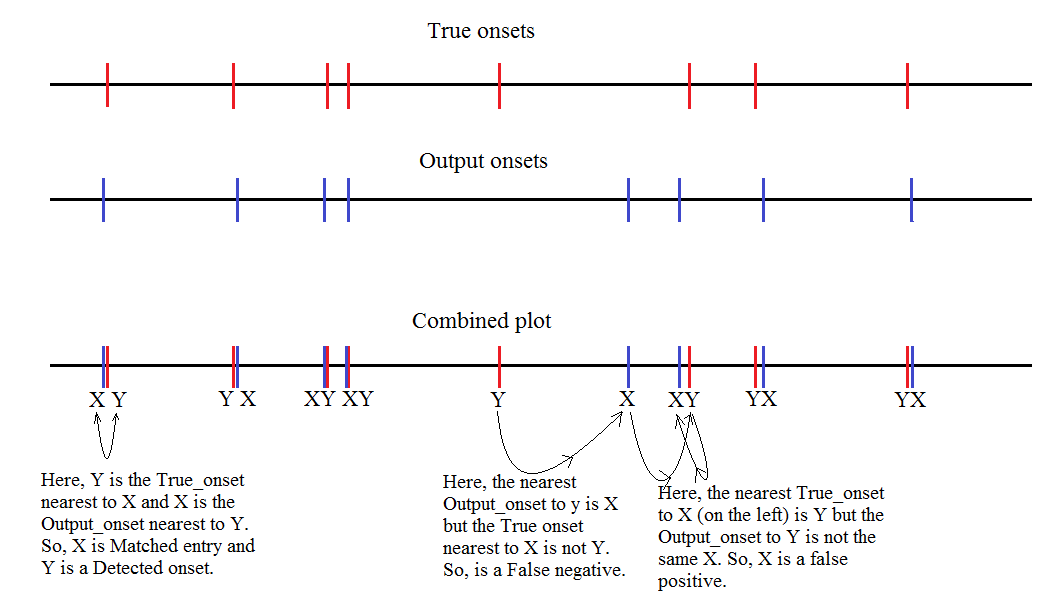}
    \caption{The Steps of Subset Matching Algorithm}
    \label{fig:subset_matching}
\end{figure}

Note that, if the Output\_onsets and True\_onsets are sorted in increasing order of time, we first perform a Merging step like in merge sort to sort both of them together. In that way, the steps for computing the nearest onsets $X$ and $Y$ can be done in a constant amount of time, just by considering the immediate neighbouring onsets to its both sides. The idea of subset matching is described in Figure \ref{fig:subset_matching}.

\subsection{Correlative Matching}

\subsubsection{The Problem}
In subset matching we assumed that the hummed song and the song present on the database are in the same time. But in reality that does not occur. The song present in the database and the hummed song may have different tempos. For example, our national anthem "Jana Gana Mana" should be sung in exactly 52 seconds, however, one might not be so precise when singing it naturally, hence, he (or she) might end up singing the anthem in 45 seconds. In such case, the basic assumption of Subset matching fails. We extend the idea subset matching for these case and call the resulting algorithm Correlative matching.

The input to this algorithm is the same as the Subset matching. However, along with the output of Matched\footnote{Here, matching means in exact sense of the song, in different scaling of time units, not in the absolute sense of time.} Output\_onsets and True\_onsets, we output a similarity score which determines how likely it is that True\_onsets and Output\_onsets, both are of same song.

\subsubsection{A Naive Approach}
Suppose we have $n$ length vector of Output\_onsets and $m$ length vector of True\_onsets. For now, assume that there is no False negatives that is all the True\_onsets are detected. So we have to choose $m$ onsets from Output\_onsets which correspond to the True\_onsets. A naive approach will be to consider all the $m$ subsets of Output\_onsets and calculate their correlation with True\_onsets. The subset with the maximum correlation will be our Predicted onsets. The complexity of this algorithm is $O\left(\binom{n}{m}k\right) \approx O(n^mk)$ where $O(k)$ is the complexity of the subset matching algorithm. The main problem of this approach is the demand of extreme need of computational time. This could lead us to a problem as the number of songs in the database can be huge and the time required for searching would go up by a significant factor. 

\subsubsection{Idea of improvement over naive approach}
To find the best matching subset we first need to change the location and scale of our True\_onsets so that it matches with our Output\_onsets. To choose the location and scale, consider that the first True\_onset corresponds to one of the first $(n-m+1)$ Output\_onsets and the last True\_onset corresponds to one of the last $(n-m+1)$ Output\_onsets. So, for all the $(n-m+1)^2$ choices of the first and last True\_onsets we obtain a linear transformation and apply that to our True\_onsets. Then for each of the choices we calculate the correlation coefficient between True\_onsets and the output of subset matching algorithm applied on Output\_onsets and transformed True\_onsets. We then multiply a correction factor to the obtained correlation to penalize for each false positives and false negatives. We output the subset that produces the maximum similarity coefficient with True\_onsets.

\subsubsection{Algorithm}
The Algorithm is described in the following points;
\begin{enumerate}
    \item Consider cor\_mat as the $(n-m+1)\times(n-m+1)$ matrix which will store the correlation coefficients between True\_onsets and the output of subset matching applied on appropriately transformed Output\_onsets.
    \item FOR i = 1 to (n-m+1):\\
    FOR j = m to n:
    \begin{enumerate}
        \item $\beta$ = (Output\_onset[j]-Output\_onset[i])/(True\_onset[m]-True\_onset[1])
        \item $\alpha$ = Output\_onset[i]-$\beta\times$True\_onset[1]
        \item Scaled\_True\_onset = $\alpha$ + $\beta\times$True\_onset. Note that, this contains the transformed True\_onsets which are in same units of time as Output\_onsets.
        \item Let $a$ = Matched entries of Subset Matching Algorithm (Output\_onset, Scaled\_True\_onset), $b$ = Detected onsets of Subset Matching Algorithm (Output\_onset, Scaled\_True\_onset)
        \item Let $L$ = Length of $a$ = Length of $b$.
        \item Number of False Positives, $FP$ = $n - L$ 
	    \item Number of False Negatives, $FN$ = $m - L$
	    \item Correction Factor = $(1 - \dfrac{FP}{n})(1 - \dfrac{FN}{m}) = \dfrac{L^2}{mn}$. Correction Factor introduces a penalty that depends linearly on the number of False positives and False negatives.
	    \item Let, cor\_mat[$i$,$j-m+1$] = cor(a, b) * Correction Factor, where $cor(\cdot)$ denotes the Pearson's product moment correlation coefficient.
    \end{enumerate}
    \item Find $i$ and $j$ for which cor\_mat[$i$,$j-m+1$] is maximum. Let $$(i^*, j^*) = \arg\max_{(i,j)}\text{cor\_mat}[i,j-m+1]$$
    \item Find the Matched entries corresponding to this choice of $i^*$ and $j^*$ just as in step 2.
    \item Return Matched\_entries as “Predicted onsets” and cor\_mat[$i^*$,$j^*-m+1$] as “Similarity coefficient”.
\end{enumerate}

Now that we have an algorithm to calculate “Similarity coefficient” between the hummed song and any song in the database, all that is left is to calculate the similarity coefficient between the hummed song and all the songs in the database. We return the song with the maximum similarity with the hummed song. 
Note that, if we have two or more songs that have similarity coefficient very close to that of the maximum similarity then we return those songs in order of similarity. The definition of closeness depends on how many songs we want to return.

\section{Experimental Results}
\qquad We perform the above algorithms with 3 hummed songs and with a database of 10 songs for the purpose of comparison between different techniques and also finding out the effectiveness of the system as a whole. The database contains the onsets of the first line of each of the 10 songs, obtained from the instrumental notations for those songs. The 3 hummed songs are \textit{Sa Re Jahan Se Accha}, \textit{Ekla Cholo Re} and \textit{Jingle Bells}. Along with these songs in the database, we put another 7 songs in the database of similar genre, and of each of 3 languages (Hindi, Bengali and English).  

The study can be separated into three major parts.

For the first part, we obtain the actual onsets from the instrumental notations of the song. Also, we visualize the oscillogram of the hummed song in \textit{Audacity}, and manually obtain the true values of the onsets of the hummed song, by merely inspecting the amplitude envelope. Then, we compute the correlation coefficient between these two sets of onsets, in order to ensure that the idea of onset is extremely prominent to every individual (even if he/she cannot sing well), and this idea can be actually used to search songs.

For the second part, we apply different detection functions and evaluate their comparative performances. This step also allows us to obtain the optimal parameter setting for different onset detection algorithms including the peak detection procedure.

Finally, the searching algorithm is performed in order to assess the quality of the system as a whole. 

The corresponding code and datasets for these experiments is available in the following \textit{Github} repository; \url{https://github.com/subroy13/OnsetDetection}. All the codes have been written in \textit{R} programming language. 

\subsection{First Song}
\qquad For the first song \textit{Sa Re Jahan Se Accha}, it is hummed in a way so that the onsets are stressed by the singer. For this reason, we expect the results should be good. This hummed song is mainly used to obtain the optimal parameter setting for different choices of detection function.

\begin{figure}
    \centering
    \includegraphics[width = \textwidth]{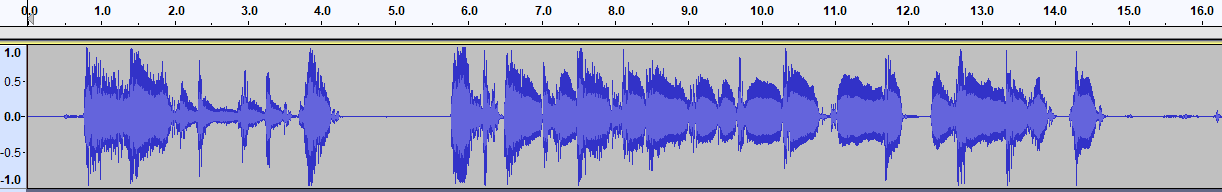}
    \caption{Oscillogram of the hummed song of Sa Re Jahan Se Accha}
    \label{fig:oscillo_song1}
\end{figure}

We obtain the true onsets of the hummed version of the song by manually visualizing the oscillogram in \textit{Audacity} which is shown in Figure \ref{fig:oscillo_song1}. The correlation of these inspected onsets with that of onsets in instrumental notations, is found to be $0.9992221$, which is extremely close to 1.  

\begin{figure}
    \centering
    \includegraphics[width = \textwidth]{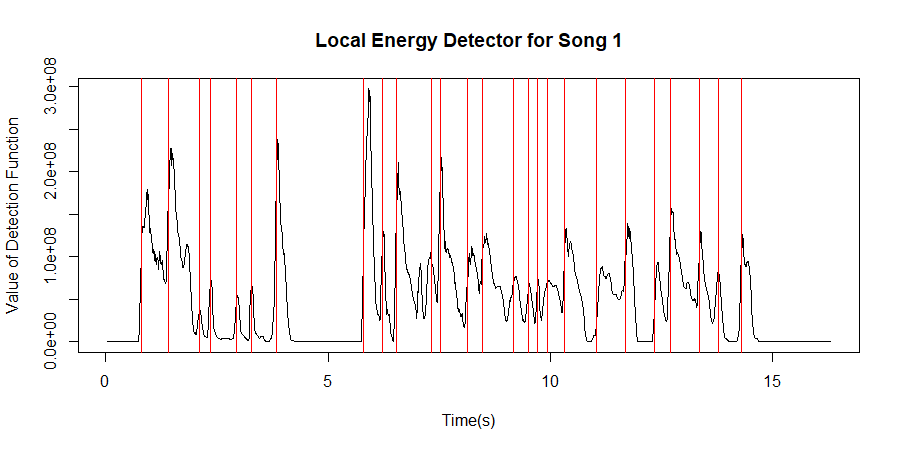}
    \caption{Values of Local Energy detection function for first song (Red lines show the times of true onsets)}
    \label{fig:energy_song1}
\end{figure}

We first apply the Local energy detector as our detection function. We apply it on a window length of $\omega = 4096$. The hummed song has a sampling frequency of $48000$ samples per second, thereby about $4800$ many samples in $0.1$ seconds. The choice of the window length is made based on the maximum power of $2$\footnote{This window length is also used in STDFT where having window length as power of 2 speeds up the computation using techniques of Fast Fourier Transformation (FFT)} such that the corresponding length does not go beyond $0.1$ seconds, the reason explained as before. The values of the detection function along with the manually obtained true peak values are given in Figure \ref{fig:energy_song1}. As we can see from the figure, the detection function is not foolproof, and there is a peak without any true onset at about 7 seconds which probably will be identified as a false positive. However, we see that there is no evident false negative in the detection function. We also see the evidence of relatively smaller size peaks at the time of true onsets.

\begin{figure}
    \centering
    \includegraphics[width = \textwidth]{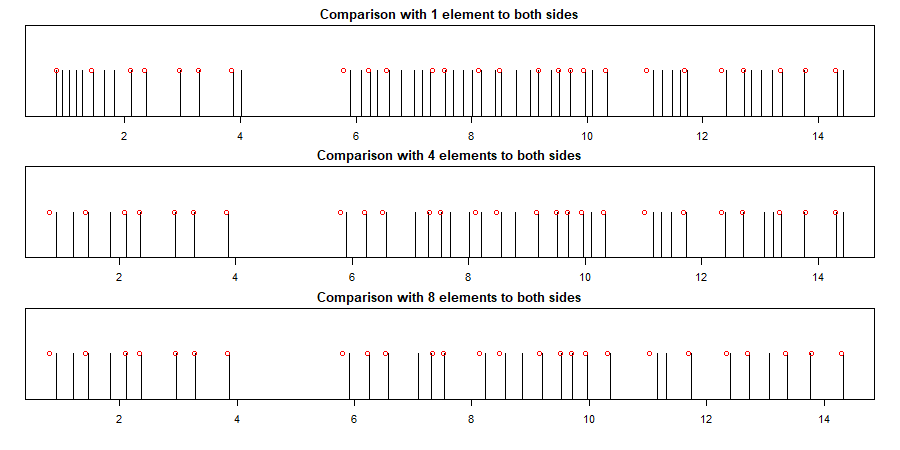}
    \caption{Obtained Peaks from Energy Detector for different parameters with mean based thresholding (Red circles show true onsets)}
    \label{fig:energy_onsets_song1}
\end{figure}

To identify the peaks in this local energy detection function values obtained, we use three different extremes of neighbouring points, respectively by choosing 1 samples, or 4 samples or 8 samples (in units of hopsize) to both sides as neighbouring points, based on a hop size of $512$ samples. We also use a mean based thresholding criterion in this case. The outputs are graphically summarized in Figure \ref{fig:energy_onsets_song1}. It is found that using 8 points in both sides as neighbouring points for comparison during peak detection is a good idea.

\begin{figure}
    \centering
    \includegraphics[width = \textwidth]{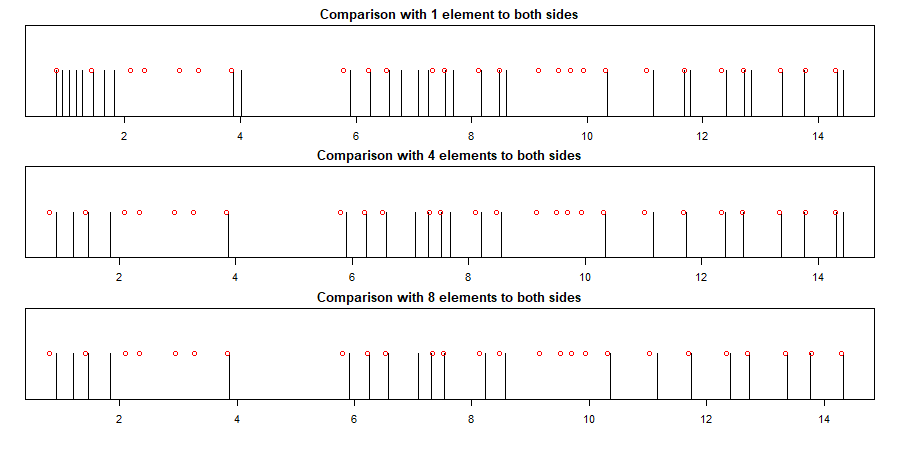}
    \caption{Obtained Peaks from Energy Detector for different parameters with Quartile based thresholding (Red circles show true onsets)}
    \label{fig:energy_onsets_song1_type2}
\end{figure}

Since the local energy detection function does not stay 0 most of the time, therefore, it performs poorly if we use quartile thresholding, as shown in Figure \ref{fig:energy_onsets_song1_type2}. Therefore, we choose to use mean based thresholding for Local energy detector for future usages.

\begin{figure}
    \centering
    \includegraphics[width = \textwidth]{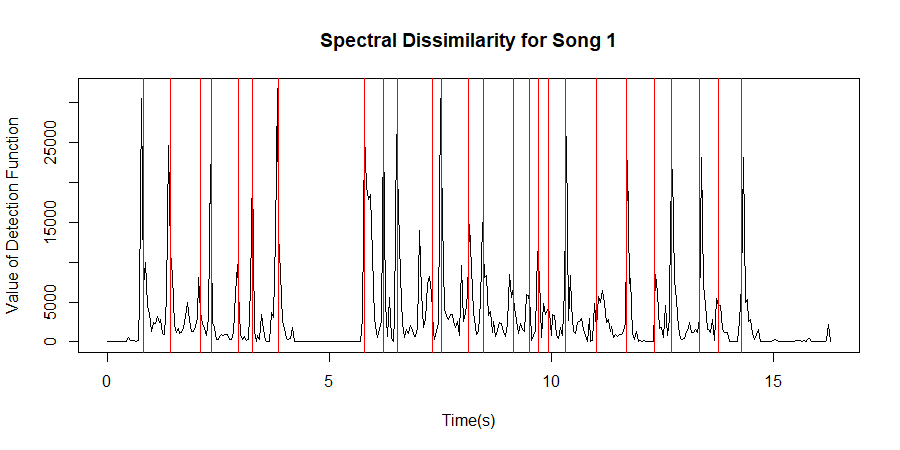}
    \caption{Values of the Spectral Dissimilarity Function for first song (Red lines show the time of true onsets)}
    \label{fig:spectral_song1}
\end{figure}

\begin{figure}
    \centering
    \includegraphics[width = \textwidth]{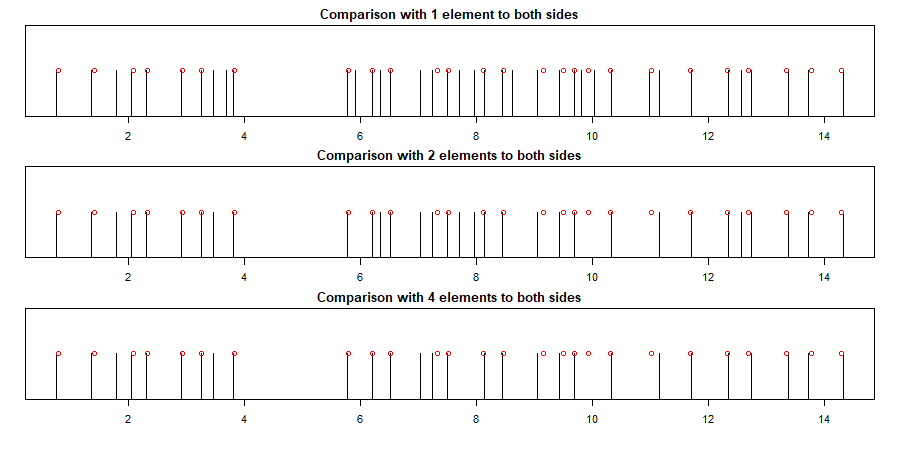}
    \caption{Obtained Peaks from Spectral Dissimilarity Detector for different parameters with mean based thresholding (Red circles show true onsets)}
    \label{fig:spectral_onsets_song1}
\end{figure}

\begin{figure}
    \centering
    \includegraphics[width = \textwidth]{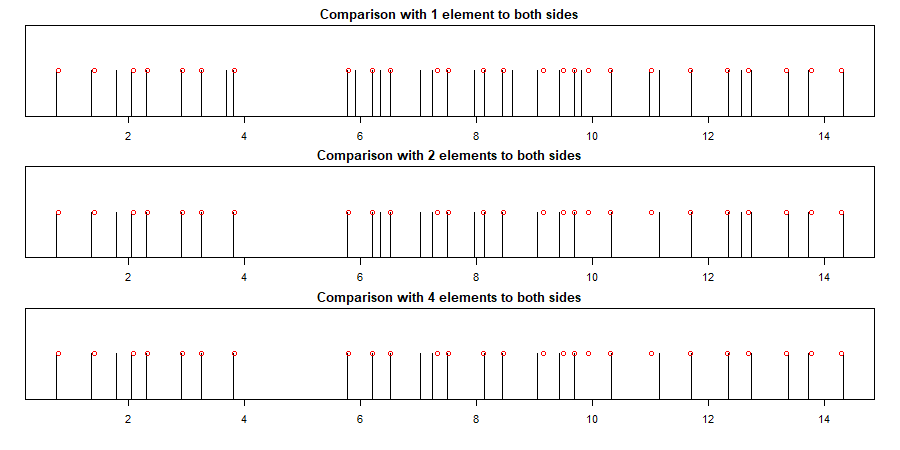}
    \caption{Obtained Peaks from Spectral Dissimilarity Detector for different parameters with quartile based thresholding (Red circles show true onsets)}
    \label{fig:spectral_onsets_song1_type2}
\end{figure}

We perform the above analysis for Spectral Dissimilarity also. Figure \ref{fig:spectral_song1} shows the values of the Spectral Dissimilarity function obtained for window length of $\omega = 4096$ samples, for reasons mentioned earlier. Note that, the hopsize for this detection algorithm should be set to high values ($2048$ samples in our study), as the changes in the function is very rapid and we do not want to detect any unnecessary changes in spectrum. Also, the effect of the choice of neighbouring points with mean based and quartile based thresholding is shown in Figure \ref{fig:spectral_onsets_song1} and Figure \ref{fig:spectral_onsets_song1_type2} respectively. We decide to use 4 points to both sides for comparison in peak detection algorithm. It also seems that for this detection function, the mean thresholding and median thresholding both performs equally well. However, for our study, we decide to use the mean thresholding. 

\begin{figure}
    \centering
    \includegraphics[width = \textwidth]{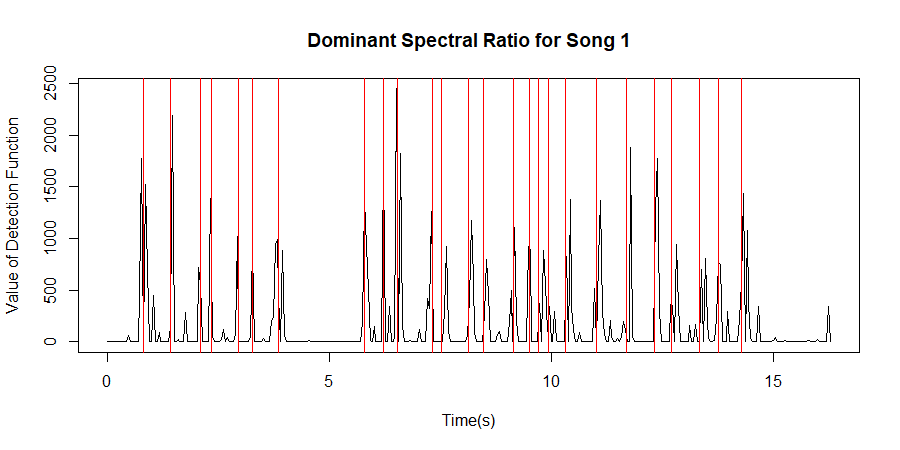}
    \caption{Values of the Dominant Spectral Dissimilarity Function for first song (Red lines show the time of true onsets)}
    \label{fig:dfreq_song1}
\end{figure}

\begin{figure}
    \centering
    \includegraphics[width = \textwidth]{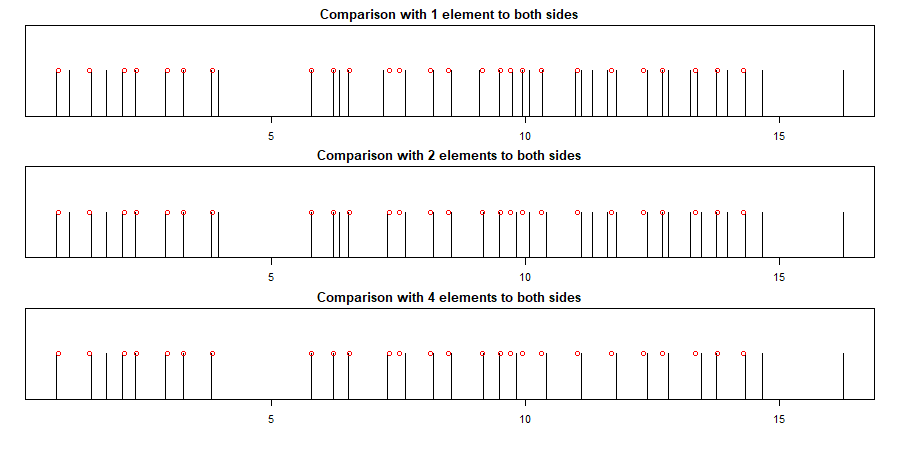}
    \caption{Obtained Peaks from Dominant Spectral Dissimilarity Detector for different parameters with mean based thresholding (Red circles show true onsets)}
    \label{fig:dfreq_onsets_song1}
\end{figure}

\begin{figure}
    \centering
    \includegraphics[width = \textwidth]{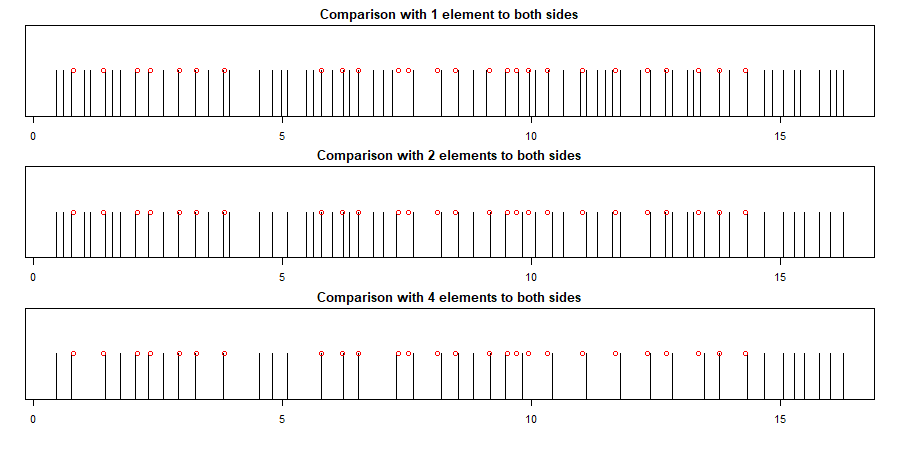}
    \caption{Obtained Peaks from Dominant Spectral Dissimilarity Detector for different parameters with quartile based thresholding (Red circles show true onsets)}
    \label{fig:dfreq_onsets_song1_type2}
\end{figure}

For Dominant Spectral Dissimilarity detection function with the same window length $\omega = 4096$ and hopsize being $2048$, we also perform a similar exercise. Figure \ref{fig:dfreq_song1} shows different values of detection function for the first hummed song. Again, it is seen that the onsets are reflected as much higher peaks in the detection function. Figure \ref{fig:dfreq_onsets_song1} and \ref{fig:dfreq_onsets_song1_type2} summarizes the performance of the detector under various different neighbour setups along with different thresholding criterion. From the outputs, it seems more reasonable to work with 2 neighbours on both sides, with mean based thresholding criterion. It is also possible to work with 4 neighbouring values to both sides, however, there is an undetected onset with such parameter setup, suggesting it slight inability to detect two very close peaks. Therefore, we stick with 2 neighbours comparison rule with this detection function.

To summarize, we choose to work with the following three detectors at their seemingly optimal parameter setting;

\begin{enumerate}
    \item Local Energy Detector, with a window length of $4096$ samples, having an hopsize of $512$ samples. For peak detection procedure, it computes a peak with its 8 neighbouring points to both sides, and use a mean based thresholding criterion.
     \item Spectral Dissimilarity Detector, with a window length of $4096$ samples, having an hopsize of $2048$ samples. For peak detection procedure, it computes a peak with its 4 neighbouring points to both sides, and use a mean based thresholding criterion.
    \item Dominant Spectral Dissimilarity Detector, with a window length of $4096$ samples, having an hopsize of $2048$ samples. For peak detection procedure, it computes a peak with its 2 neighbouring points to both sides, and use a mean based thresholding criterion.
\end{enumerate}

Finally, we apply our searching techniques to this obtained onsets in order to identify the song. Table \ref{tbl:search_song1} summarizes the output of the Searching algorithm performed on the database. It shows the top 5 songs that matches most to the hummed song along with their corresponding scores. We find that, the searching algorithm always finds the correct song whichever detection algorithm is performed beforehand.

\begin{table}
\centering
\caption{Details of Searching Output using hummed version of \textit{Sa Re Jahan Se Accha}}
\label{tbl:search_song1}
\resizebox{\textwidth}{!}{
\begin{tabular}{|c|c|c|c|c|c|c|}
\hline
\multirow{2}{*}{Ranks} & \multicolumn{2}{c|}{Energy Detector} & \multicolumn{2}{c|}{Spectral Dissimilarity} & \multicolumn{2}{c|}{Dominant SD} \\ \cline{2-7} 
                       & Song                     & Score     & Song                         & Score        & Song                   & Score   \\ \hline
1                      & Sa Re Jahan Se Accha     & 0.75      & Sa Re Jahan Se Accha         & 0.791        & Sa Re Jahan Se Accha   & 0.649   \\ \hline
2                      & Jingle Bells             & 0.573     & Jingle Bells                 & 0.595        & Fur Elise              & 0.623   \\ \hline
3                      & Jana Gana Mana           & 0.568     & Jana Gana Mana               & 0.585        & Jingle Bells           & 0.544   \\ \hline
4                      & Hain Apna Dil            & 0.53      & Jindegi Ek Safar             & 0.578        & Jana Gana Mana         & 0.541   \\ \hline
5                      & Jindegi Ek Safar         & 0.506     & My Heart will go on          & 0.578        & Ore Grihobasi          & 0.493   \\ \hline
\end{tabular}
}\end{table}

\subsection{Second Song}
\qquad The second song chosen to be analyzed is \textit{Ekla Cholo Re}. It is a popular Rabindra Sangeet with meends in its notes. This hummed song kind of depicts the drawbacks of Local energy detector, and lays out the important idea why rather than the absolute value of the amplitude, the detection function should be able to convey the absolute change in amplitude envelopes.

\begin{figure}
    \centering
    \includegraphics[width = \textwidth]{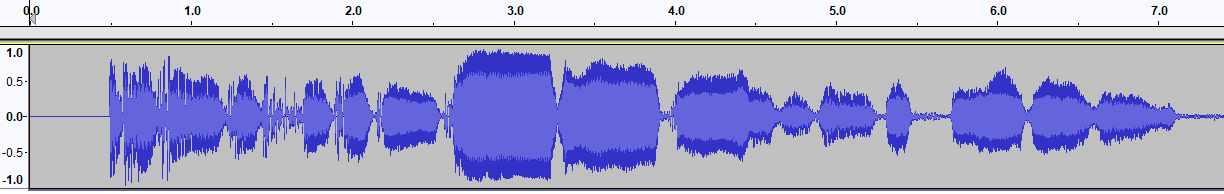}
    \caption{Oscillogram of the hummed song of \textit{Ekla Cholo Re}}
    \label{fig:oscillo_song2}
\end{figure}

We obtain the true onsets of the hummed version of the song by manual inspection of the amplitude profile in \textit{Audacity}, as shown in Figure \ref{fig:oscillo_song2}. The original intrumental onsets have a correlation of $0.9998769$ with the hummed version of the song.

\begin{figure}
    \centering
    \includegraphics[width = \textwidth]{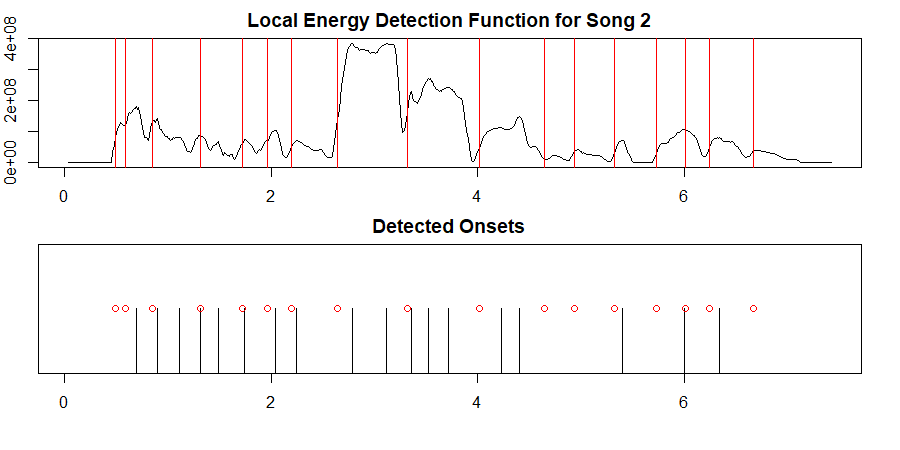}
    \caption{Performance of Local Energy Detector for \textit{Ekla Cholo Re}}
    \label{fig:energy_song2}
\end{figure}

Based on the optimal setting for Local Energy detector, we compute the detection function and obtain the peaks. Figure \ref{fig:energy_song2} shows how bad the performance of Local Energy Detector is affected due to the presence of meends. As we see, there are lots of false negatives in the detected onsets, which should cause a detrimental effect to the performance of the searching algorithm. However, as seen from Table \ref{tbl:search_song2}, Local energy detector correctly identify the song even in this case. Based on the insight of the searching algorithm, we think the only possible reason of this is because the size of the database is extremely small. However, had the size of the database been bigger, the system would fail to identify the song correctly.

\begin{figure}
    \centering
    \includegraphics[width = \textwidth]{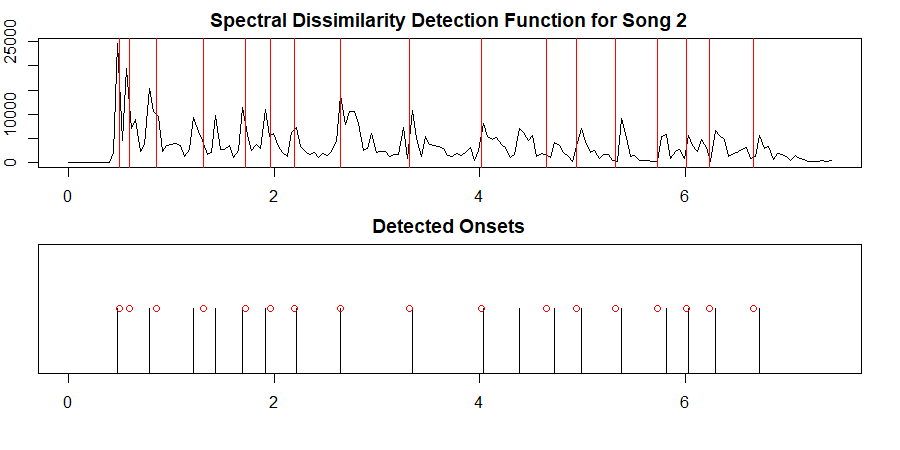}
    \caption{Performance of Spectral Dissimilarity Detector for \textit{Ekla Cholo Re}}
    \label{fig:spectral_song2}
\end{figure}

Applying Spectral Dissimilarity Detection function with its optimal setting, we obtain a much better values of detected onsets, as seen from Figure \ref{fig:spectral_song2}. Also note that, there is one possible false negative during the starting of the song. The occurrence of two true onsets being too close to each other might be the reason of such undetected onsets, one of which is dominated by the other during the comparison for Peak Detection procedure. From Table \ref{tbl:search_song2}, we see that, based on these detected onset timepoints, our searching algorithm finds the correct song as its first preference. We also find that the difference of scores in first and second choice are about $0.18$, signifying the amount of certainty.

\begin{figure}
    \centering
    \includegraphics[width = \textwidth]{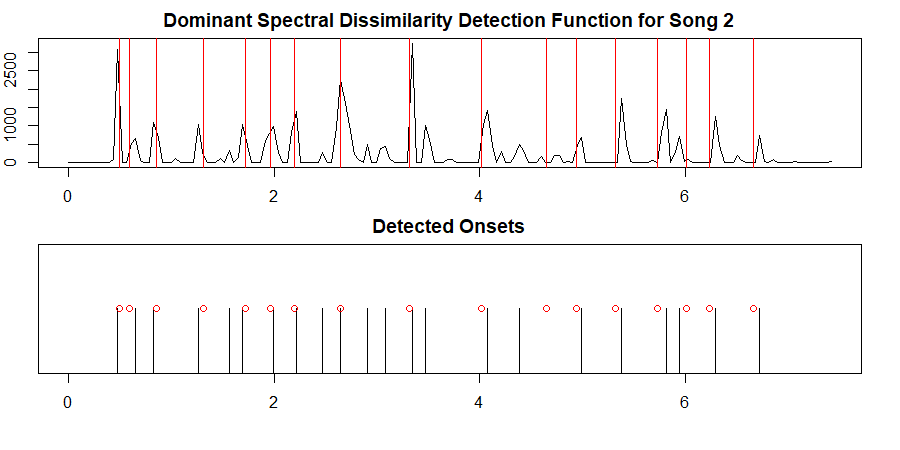}
    \caption{Performance of Dominant Spectral Dissimilarity Detector for \textit{Ekla Cholo Re}}
    \label{fig:dfreq_song2}
\end{figure}

Finally, we apply Dominant Spectral Dissimilarity detection function on this hummed song at its optimal hyperparameter setups. As seen from Figure \ref{fig:dfreq_song2}, this algorithm performs similar to the Spectral Dissimilarity detection function. Note that, the onset that Spectral Dissimilarity failed to detect, has been detected by this algorithm, however, paying a cost of missed detection of another onset. However, also note that, this detection function comes with many false positives, which is not there for Spectral Dissimilarity detection function. Table \ref{tbl:search_song2} shows that these detected onsets also identify the song correctly. However note that, the songs detected as 2nd or 3rd preference do not have an extremely less score than the true song. The possible reason behind this would be the detection of many false positives, which matches some onsets of other songs.

\begin{table}
\centering
\caption{Details of Searching Output using hummed version of \textit{Ekla Cholo Re}}
\label{tbl:search_song2}
\begin{tabular}{|c|c|c|c|c|c|c|}
\hline
\multirow{2}{*}{Ranks} & \multicolumn{2}{c|}{Energy Detector} & \multicolumn{2}{c|}{Spectral Dissimilarity} & \multicolumn{2}{c|}{Dominant SD} \\ \cline{2-7} 
                       & Song                  & Score        & Song                      & Score           & Song                   & Score     \\ \hline
1                      & Ekla Cholo Re         & 0.64         & Ekla Cholo Re             & 0.735           & Ekla Cholo Re          & 0.684     \\ \hline
2                      & Hain Apna Dil         & 0.552        & Hain Apna Dil             & 0.551           & Jana Gana Mana         & 0.668     \\ \hline
3                      & Jana Gana Mana        & 0.522        & Jana Gana Mana            & 0.509           & My Heart will go on    & 0.657     \\ \hline
\end{tabular}
\end{table}

\subsection{Third Song}
\qquad The third song chosen to be analyzed is the Christmas song \textit{Jingle Bells}. This song is hummed in relatively higher frequency that the previous ones, also with alternating high and low pitches. In this direction, our aim with this hummed song is to see whether the performance of the whole system reduces due to different variations of frequency and pitch of the song.

\begin{figure}
    \centering
    \includegraphics[width = \textwidth]{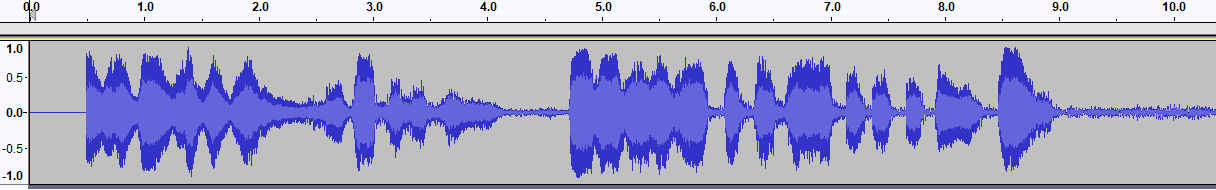}
    \caption{Oscillogram of the hummed song of \textit{Jingle Bells}}
    \label{fig:oscillo_song3}
\end{figure}

Before proceeding with detection of the onset, we obtain the true values of the onsets of the hummed song through manual inspection, as shown in Figure \ref{fig:oscillo_song3}. The correlation coefficient between these manually obtained onsets and the original onsets in the instrumental notations of the song is found to be $0.9992887$, which is again pretty close to 1.

\begin{figure}
    \centering
    \includegraphics[width = \textwidth]{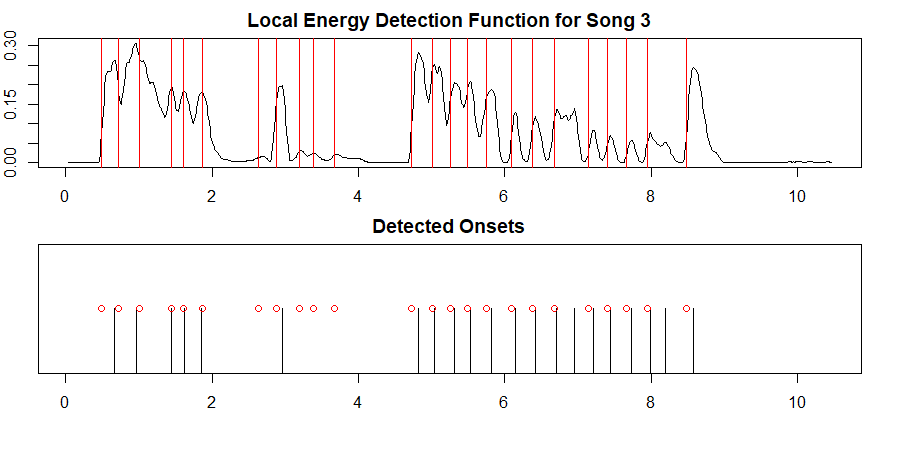}
    \caption{Performance of Local Energy Detector for \textit{Jingle Bells}}
    \label{fig:energy_song3}
\end{figure}

We perform Local energy detection function on this hummed version of the song. From Figure \ref{fig:energy_song3}, it is evident that many true onsets remain undetected by the detection algorithm. Note that, as Table \ref{tbl:search_song3} shows, this poor performance of the detection function affects the searching algorithm, which fails to identify the song correctly.

\begin{figure}
    \centering
    \includegraphics[width = \textwidth]{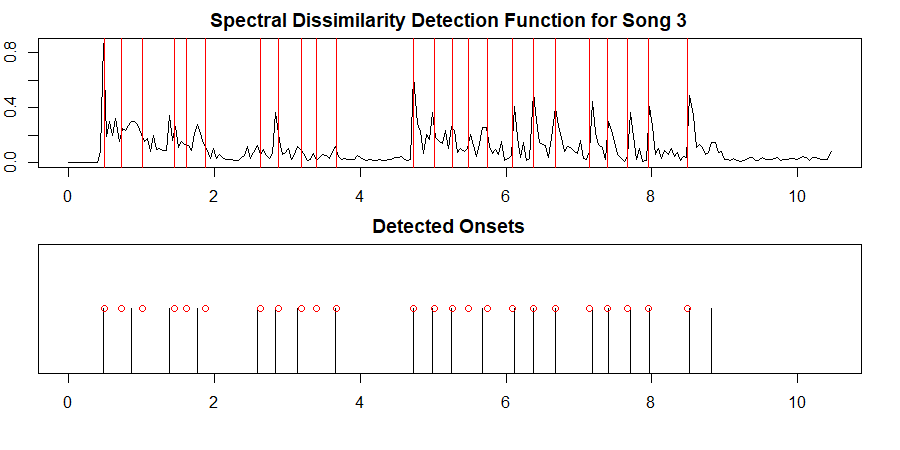}
    \caption{Performance of Spectral Dissimilarity Detector for \textit{Jingle Bells}}
    \label{fig:spectral_song3}
\end{figure}

From the detected onsets based on Spectral Dissimilarity detection function, as seen from Figure \ref{fig:spectral_song3}, it seems that the problem with missed detection of onsets still persists. Table \ref{tbl:search_song3} also shows the inability to identify the song correctly in this case. However, note that the song \textit{Jingle Bells} has same repeating pattern of onsets at the very beginning as \textit{Jindegi Ek Safar} which is obtained as the first preference by the searching algorithm. The inability of detecting the true onsets is possibly due to the sudden drop in pitch by the singer of the song. Also, we see that the optimal criterion of selecting 4 neighbours for comparison in case of peak detection procedure, may not be good in this case.

\begin{figure}
    \centering
    \includegraphics[width = \textwidth]{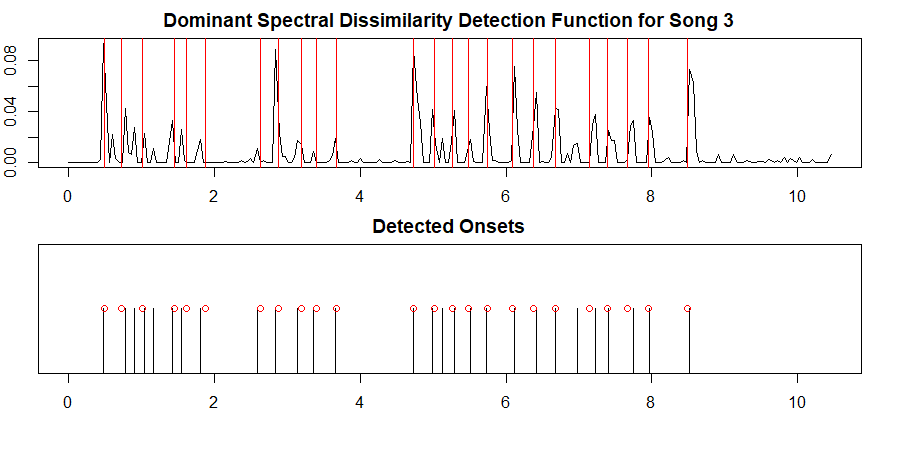}
    \caption{Performance of Dominant Spectral Dissimilarity Detector for \textit{Jingle Bells}}
    \label{fig:dfreq_song3}
\end{figure}

Finally, Dominant Spectral Dissimilarity detection function shows a more promising output as seen in Figure \ref{fig:dfreq_song3}. Note that, although there are some false positives outputted by the algorithm, it identifies all of the true positives. From Table \ref{tbl:search_song3}, we see that this detection function outputs the correct song via the searching algorithm. On this note, it is reasonable to say that having an undetected onset is more harmful than detecting a false onset, for our specific algorithm. Therefore, it is advised to tune the peak detection algorithm in a way so that it compares less number of neighbours to both sides when detecting peaks.

\begin{table}
\centering
\caption{Details of Searching Output using hummed version of \textit{Jingle Bells}}
\label{tbl:search_song3}
\begin{tabular}{|c|c|c|c|c|c|c|}
\hline
\multirow{2}{*}{Ranks} & \multicolumn{2}{c|}{Energy Detector} & \multicolumn{2}{c|}{Spectral Dissimilarity} & \multicolumn{2}{c|}{Dominant SD} \\ \cline{2-7} 
    & Song                  & Score        & Song                      & Score           & Song                   & Score     \\ \hline
1   & Jindegi Ek Safar  & 0.608  & Hain Apna Dil & 0.716  &  Jingle Bells    & 0.72     \\ \hline
2       & Hain Apna Dil         & 0.549        & Jindegi Ek Safar             & 0.716           & Jindegi Ek Safar         & 0.644     \\ \hline
3   &   Ekla Cholo Re     & 0.548        & Jingle Bells   & 0.687           & Hain Apna Dil   & 0.607  \\ \hline
\end{tabular}
\end{table}

\subsection{Power Calculation}
\qquad In our experiments, all three hummed songs were recorded at a sampling frequency of 48000 samples per second. Also, we record a blank audio (without any onset) of 1 second which helps us to estimate the usual noise variance $\sigma^2$. Based on the normal model we have introduced earlier, we find the estimated noise variance to be $10165.98$ units. Also, for each song, we compute its mean energy level i.e. $T = \sum_{i=0}^{N-1}x^2[i]/N$, and estimate the squared signal to noise ratio (SSNR) of that song as $T/10165.98$, which is a very crude underestimate, as SNR is defined to be $A/\sigma$, where $A$ is the highest amplitude obtained at the time of the onset, which is surely greater than or equal to the average amplitude of the whole song. We find that the sqaured value of SNR is roughly estimated as $5000$ for each of the song, hence we use this to calculate the power of the energy detector.

\begin{figure}
    \centering
    \includegraphics[width = \textwidth]{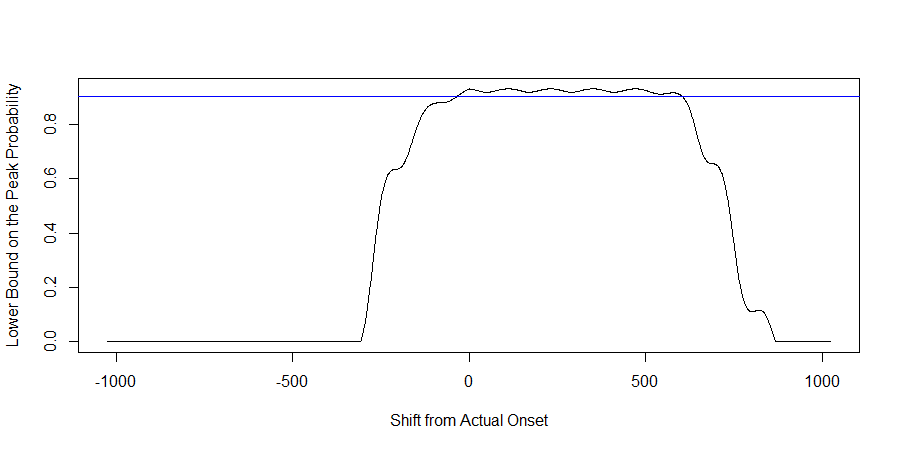}
    \caption{Lower Bound to the Probability of Peak Detection (Blue line shows the 90\% assurance line)}
    \label{fig:power_energy}
\end{figure}

From the Figure \ref{fig:power_energy}, it seems evident that Local Energy Detector, under the assumption of the model, detects the onset with very high probability (at least 0.9) in the region between about $100$ samples before the true onset to about $650$ samples after the true onset. Therefore, if we choose a hopsize of $512$ samples, then it is sure that we run the peak detection algorithm at least once within the above described interval, thus detecting the true onset with at least 90\% chance.

On the other hand, with window length $\omega = 4096$, and Squared SNR as $5000$, the mean threshold $\alpha$ can be estimated as $5000\times \hat{\sigma}^2$, where $\hat{\sigma^2}$ is the estimate of noise variance. Therefore, $P(T[k] > \alpha) = P(T[k]/\sigma^2 > \alpha/\sigma^2)$, here, $\alpha/\sigma^2$ can be estimated using the squared signal to noise ratio (i.e. by $5000$). Therefore, approximately, the probability of having a false positive at least $4096$ samples away from the true onset, is $4.704\times 10^{-21}$, which is extremely small. 

\begin{figure}
    \centering
    \includegraphics[width = \textwidth]{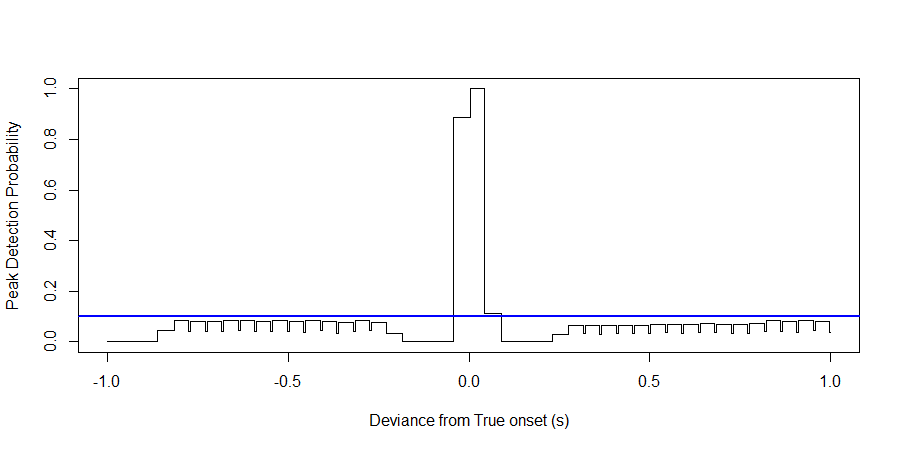}
    \caption{Estimated Probability of Peak Detection for Spectral Dissimilarity (Blue line denotes 0.1 probability)}
    \label{fig:power_spectral}
\end{figure}

\begin{figure}
    \centering
    \includegraphics[width = \textwidth]{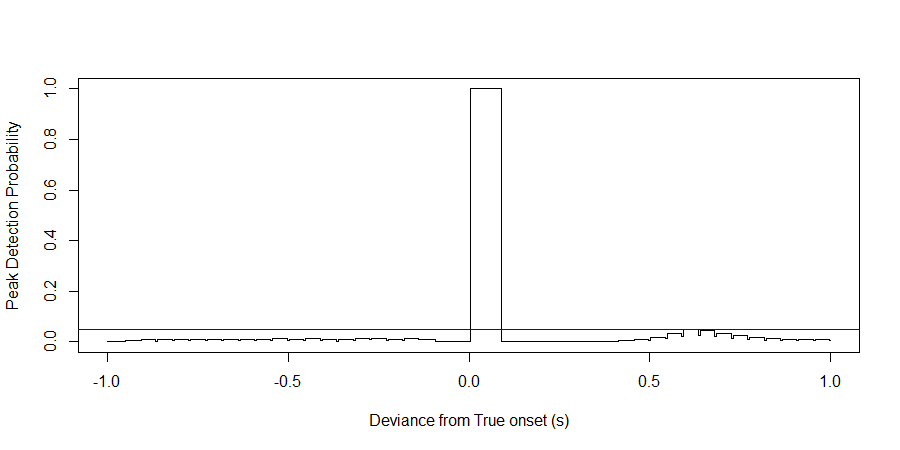}
    \caption{Estimated Probability of Peak Detection for Dominant SD (Blue line denotes 0.05 probability)}
    \label{fig:power_dfreq}
\end{figure}

Since, the distributions of the test statistics of Spectral Dissimilarity and Dominant Spectral Dissimilarity is difficult to obtain in closed form, we tried to simulate data from our assumed model and apply the detection algorithm in order to evaluate its performance. We repeat this process $10,000$ times, and obtain the proportion of times the detection algorithm outputs an onset at some position out of $10,000$ trials. Figure \ref{fig:power_spectral} shows that with very high probability, an onset should be detected within 0.05 seconds margin of the true onset. Also, there is less than 10\% chance of a noise being identified as an onset, which is the bound for the probability of false positive produced by the algorithm.

Figure \ref{fig:power_dfreq} shows the same obtained for Dominant Spectral Dissimilarity detection function. It seems that this algorithm does not detect an onset before the value of the true onsets, hence in a sense, the estimated onset will be a biased with positive bias. However, the probability of false positive seems to be reasonably low, less than $0.05$. In this sense, this algorithm should perform better than Spectral Dissimilarity when the model assumptions are valid.

In this regard, it is worth noting that when many instruments are being played simultaneously (like an orchestra), i.e. for every onset, there are many frequencies for which the frequency content increases rapidly, then Dominant Spectral Dissimilarity detection function should work poorly than Spectral Dissimilarity, as the former does not use the changes in whole range of frequency spectrum.

\section{Conclusion}
\qquad From the above analysis and experimental results, it seems evident that our goal of searching the song based on the humming is fulfilled in some extent. We obtained the optimal parameter setup for all three algorithms and noted that it is best to tune the parameter in a way to assure the nonexistence of any undetected onsets with high probability. We also found that the Local Energy detector does not perform well under different note variants (or Alankars, e.g. Meend), and Spectral Dissimilarity detection suffers from the changes in the noise pattern. In this regard, Dominant Spectral Dissimilarity is better than the previous two approaches, as it does not skip a true onset. Although, it suffers from the detection of more false positives than other approaches, the system as a whole works better with this.

\qquad Our approach to perform QBH using the onsets of the hummed song has three main advantages over the conventional methods.
\begin{enumerate}
    \item Most of the typical methods for QBH use the pitch information in the hummed song. Our method relies more on the rhythm of the song. So, when the input is somewhat inharmonious or off-key most of the traditional systems fail but our method performs rather well.
    \item Our method compares the onsets of the hummed song with that of the songs in the database. This eliminates the need of storing the entire songs in the database. We just need to store the onsets of the song in the database. This results is a huge reduction in the storage space.
    \item There is another drawback in most QBH systems. Most of them require a song to be sung by a handful of singers so that the algorithm can compare the hummed song with the different versions of the same song. This entails a lot of human effort to generate the database. It is the reason why most of the QBH systems can't be employed on a large scale. Our approach doesn't require any human singing other than the input itself. It only requires the onsets of the actual song which can be found easily from online sources or can be done by hand by an expert.
\end{enumerate}

\section{Future Scopes}
\qquad Given limited time and resources, there are lots of things that could not be done. However, we think, the following directions might serve as good future scopes of this project.

\begin{enumerate}
    \item We perform the analysis assuming the fact that the noise are independent and identically distributed according to normal distribution with mean $0$ and constant variance $\sigma^2$. However, in practice, this assumption might not be true, as there might be different correlation structure between these errors.
    \item It would be better if some tighter bounds on the probability of type I error and type II errors are available for the three detection algorithms.
    \item Usage of dynamic time warping method rather than simple correlation to improve the searching technique.
    \item Our algorithm demands the user to sing the first verse of the song. However, the point where the first verse ends may not be known to the user. In such case, searching procedure also should be able to match prefix of the song.
\end{enumerate}

\section{Acknowledgements}
\qquad We would like to thank Dr. Arnab Chakraborty for introducing the problem to us and giving us an opportunity to work on it. We would also like to extend our thanks to our classmates who helped us to build the datasets by providing us with their humming of the selected songs.

\end{document}